%% file: MSOM-template.tex
\newif\ifpreprint
\preprinttrue  

\ifpreprint

\documentclass[12pt]{article}

\title{Equitable Data-Driven Facility Location and Resource Allocation to Fight the Opioid Epidemic}
\author{Joyce Luo and Bartolomeo Stellato}

\usepackage[authoryear]{natbib}
\def\bibfont{\footnotesize}

\usepackage{amsmath,enumitem,mathtools,amsthm,amssymb}

\usepackage{fullpage}
\usepackage[titletoc,title]{appendix}
\usepackage[hyphens]{url}
\usepackage[hidelinks]{hyperref}
\usepackage{adjustbox}
\usepackage{multirow}
\allowdisplaybreaks

\else

\documentclass[msom,blindrev]{informs3} 

\DoubleSpacedXI

\usepackage{natbib}
 \bibpunct[, ]{(}{)}{,}{a}{}{,}%
 \def\bibfont{\small}%
 \defcitealias{SVI}{Centers for Disease Control 2021}
 \defcitealias{NSDUH}{U.S. Department of Health and Human Services 2019}
 \defcitealias{NSSATS}{U.S. Department of Health and Human Services 2021}
 \defcitealias{nhanes}{Centers for Disease Control 2019}

\TheoremsNumberedThrough     

\EquationsNumberedThrough    

\allowdisplaybreaks

\TITLE{Equitable Data-Driven Facility Location and Resource Allocation to Fight the Opioid Epidemic}

\ARTICLEAUTHORS{%
\AUTHOR{Joyce Luo}
\AFF{Massachusetts Institute of Technology, \EMAIL{joyceluo@mit.edu}, \URL{}}
\AUTHOR{Bartolomeo Stellato}
\AFF{Princeton University, \EMAIL{bstellato@princeton.edu}, \URL{https://stellato.io}}
} 


\fi

\input{definitions.tex}

\usepackage{booktabs} 
\usepackage{pgfplots}
\usepackage{pgfplotstable}
\pgfplotsset{compat=1.10}
\usepackage{adjustbox}
\usepackage{csvsimple}
\usepackage{caption}
\usepackage{subcaption}
\usepackage{enumerate}
\usepackage{xcolor}
\newcommand{\reviewChanges}[1]{{#1}}
\newcommand{\secondReviewChanges}[1]{{#1}}
\newcommand{\thirdReviewChanges}[1]{{#1}}
\newcommand{\fourthReviewChanges}[1]{{#1}}
\usepackage{dsfont}

\newcommand{\myabstractmsom}{\textbf{\textit{Problem definition:}} The opioid epidemic is a crisis that has plagued the United States (US) for decades. One central issue of the epidemic is inequitable access to treatment for opioid use disorder (OUD), which puts certain populations at a higher risk of opioid overdose.
\textbf{\textit{Methods:}} We integrate a predictive dynamical model and a prescriptive optimization problem to compute high-quality opioid treatment facility and treatment budget allocations for each US state.
Our predictive model is a differential equation-based epidemiological model that captures the dynamics of the opioid epidemic. We use a \secondReviewChanges{process inspired by neural ordinary differential equations}
to fit this model to opioid epidemic data for each state and obtain estimates for unknown parameters in the model. We then incorporate this epidemiological model into a corresponding mixed-integer optimization problem (MIP) that aims to minimize the number of opioid overdose deaths and the number of people with OUD.
We develop strong relaxations based on McCormick envelopes to efficiently compute approximate solutions to our MIPs that have \secondReviewChanges{a mean optimality gap of 3.99\%.}
\textbf{\textit{Results:}}
Our method provides socioeconomically equitable solutions, as it incentivizes investments in areas with higher social vulnerability (from the US Centers for Disease Control’s Social Vulnerability Index) and opioid prescribing rates.
On average, \secondReviewChanges{when allowing for overbudget solutions,} our approach decreases the number of people with OUD by 
\secondReviewChanges{$9.03\pm1.772\%$}, increases the number of people in treatment by 
\secondReviewChanges{$88.75\pm26.223\%$}, and decreases the number of opioid-related deaths by 
\secondReviewChanges{$0.58\pm0.111\%$} after 2 years compared to the baseline epidemiological model's predictions.
\textbf{\textit{Managerial implications:}}
\secondReviewChanges{Our solutions show that policy-makers should target adding treatment facilities to counties that have significantly \fourthReviewChanges{fewer} facilities than their population share and are more socially vulnerable. Furthermore, we demonstrate that our optimization approach, guided by epidemiological and socioeconomic factors, should help inform these strategic decisions, as it yields population health benefits in comparison to benchmarks based solely on population and social vulnerability.  
}

}

\newcommand{\myabstract} {The opioid epidemic is a crisis that has plagued the United States (US) for decades. One central issue of the epidemic is inequitable access to treatment for opioid use disorder (OUD), which puts certain populations at a higher risk of opioid overdose.
We integrate a predictive dynamical model and a prescriptive optimization problem to compute high-quality opioid treatment facility and treatment budget allocations for each US state.
Our predictive model is a differential equation-based epidemiological model that captures the dynamics of the opioid epidemic. We use a
\secondReviewChanges{process inspired by neural ordinary differential equations}
to fit this model to opioid epidemic data for each state and obtain estimates for unknown parameters in the model. We then incorporate this epidemiological model into a corresponding mixed-integer optimization problem (MIP) that aims to minimize the number of opioid overdose deaths and the number of people with OUD.
We develop strong relaxations based on McCormick envelopes to efficiently compute approximate solutions to our MIPs that have \secondReviewChanges{a mean optimality gap of 3.99\%.}
Our method provides socioeconomically equitable solutions, as it incentivizes investments in areas with higher social vulnerability (from the US Centers for Disease Control’s Social Vulnerability Index) and opioid prescribing rates.
On average, \secondReviewChanges{when allowing for overbudget solutions,} our approach decreases the number of people with OUD by 
\secondReviewChanges{$9.03\pm1.772\%$}, increases the number of people in treatment by 
\secondReviewChanges{$88.75\pm26.223\%$}, and decreases the number of opioid-related deaths by 
\secondReviewChanges{$0.58\pm0.111\%$} after 2 years compared to the baseline epidemiological model's predictions.
\secondReviewChanges{Our solutions show that policy-makers should target adding treatment facilities to counties that have significantly \fourthReviewChanges{fewer} facilities than their population share and are more socially vulnerable. Furthermore, we demonstrate that our optimization approach, guided by epidemiological and socioeconomic factors, should help inform these strategic decisions, as it yields population health benefits in comparison to benchmarks based solely on population and social vulnerability.}
}

\begin{document}

\ifpreprint \else
\ABSTRACT{%
\myabstractmsom
}
\fi

\date{}
\maketitle

\ifpreprint
\begin{abstract}\myabstract \end{abstract}
\fi




\section{Introduction}\label{intro}
The opioid epidemic is a foremost public health crisis within the United States (US). The epidemic has been driven by increases in prescription, illicit, and synthetic opioid use, which have in turn increased rates of opioid use disorder (OUD) and overdose deaths. According to the Centers for Disease Control (CDC), around 500,000 people have died from overdoses involving both illicit and prescription opioids from 1999 to 2019~\citep{opioid_basics}. The COVID-19 pandemic has further exacerbated the opioid epidemic, with recent data showing a spike in overdose deaths during 2020. In the period from September 2019 through August 2020, there were 88,295 predicted deaths, which is about 27\% more than the preceding 12-month period~\citep{cdcwonder,covid19}. The pandemic has brought to the forefront the need for expanded access to opioid addiction treatment services.

Currently, the main treatment for OUD is medication-assisted treatment (MAT), which has been proven to sustain patient recovery and prevent future overdoses. Methadone and buprenorphine are the two main medications approved to treat OUD~\citep{disparities}.
Although access to both drugs has expanded in the last decade, there are still major gaps in access to these treatments across the US, especially in rural areas with under-developed health infrastructures. Those seeking care are often required to travel long distances to federally-approved opioid treatment programs (OTPs) or other treatment facilities, which is another major factor that affects treatment retention~\citep{disparities}. Implementing a method that proposes more equitable and impactful treatment facility and budget allocations could help improve policy decision-making related to this issue.

In this work, we formulate an approach that provides high-quality opioid treatment facility location and treatment budget allocation decisions to address the issue of inequitable opioid treatment facility access. Our approach integrates a dynamical model of the opioid epidemic with a prescriptive mixed-integer optimization problem (MIP) for each state. We model the state-level opioid epidemic using an ODE-based epidemiological model. In order to fit the model to real world data obtained from the CDC, National Institute on Drug Abuse (NIDA), and Substance Abuse and Mental Health Service Administration (SAMHSA), we use \secondReviewChanges{a neural ODE-inspired model-fitting process}. Representing the ODE model as a neural network layer through neural ODEs helps us exploit the power of gradient descent for more efficient parameter estimation compared to zeroth order methods~\citep{chen2019neural}.

We then formulate an MIP for each state to optimize resource allocation interventions that minimize the effect of the opioid epidemic. We do this by including a discretized version of the state-level dynamical model within the constraints of the respective state's MIP and setting the objective to minimize overdose deaths and the number of people with OUD. We capture the impact of the interventions by showing how they affect a particular parameter of our discretized epidemic model in each time period. \reviewChanges{Since our MIP formulations are non-linear and cannot be efficiently solved for larger problem instances, we develop strong relaxations based on McCormick envelopes \citep{mccormick} to efficiently obtain high-quality allocation solutions with small optimality gaps.}

Our method builds recommendations about how many additional treatment facilities and how much of a limited treatment budget per time period should be allocated to each county. Our approach also incorporates information about the social vulnerability and opioid prescribing rates of each county, which is a measure of how susceptible a community is to the adverse impacts caused by external stresses on human health~\ifpreprint \citep{SVI}. \else \citepalias{SVI}. \fi 
\thirdReviewChanges{
To quantify the equitability of our solutions, we develop a measure based on the social vulnerability of each county.
In addition, we develop extensions that incorporate spatial information to model resource sharing with surrounding counties, as well as explicitly ensure robustness of our solutions. }

\subsection{Related Work}
Past computational research related to the opioid epidemic has mainly centered around modeling epidemic dynamics. This research uses compartmental models
to capture the dynamics of the opioid epidemic. The Susceptible-Infected-Recovered (SIR) model, developed by \citet{sir}, is a fundamental compartmental model used in epidemiology to simulate the spread of infectious diseases such as influenza, SARS, and most recently COVID-19~\citep{doi:10.1287/opre.2022.2306}. The SIR model uses a system of ordinary differential equations (ODEs) to model transitions between the different compartments of susceptible, infected, and recovered people within a population. A modified version of this model can be developed in regards to the opioid epidemic, as the fundamental dynamics of the opioid epidemic are similar to those of infectious diseases. Becoming addicted to opioids can be seen as analogous to being ``infected'' by a disease, and entering treatment for opioid addiction can be seen as entering recovery. Although these models are a simplification of the true dynamics of disease spread, they are very useful for assessing the impact of different interventions on the way the population compartments evolve over time.

\citet{white} detail one of the first dynamical models of opiate addiction, with a focus on heroin use. Their ODE-based compartmental model gives insight into the progression of drug users, from initiation all the way to recovery or relapse.
\citet{battista} expand on \citeauthor{white}'s compartmental model, proposing a new model based on the commonly-used Susceptible-Exposed-Infected-Recovered (SEIR) model from epidemiology. Their model specifically focuses on capturing the dynamics of the prescription opioid epidemic with four compartments: Susceptible, Prescribed, Addicted, and Rehabilitation. The transitions between each class are determined by yearly rate parameters deduced from literature or from testing ranges of parameter values~\citep{battista}. We expand upon this model and other opioid epidemic compartmental modeling work by disaggregating to state level dynamics rather than national dynamics, including more population compartments, and modifying the transition dynamics.

In addition to modeling the epidemic, there is a body of literature that uses these models to project the impact of certain policy interventions on epidemic dynamics, opioid misuse, and overdose deaths. \citet{chen} formulate a compartmental model of the US opioid epidemic to project opioid overdose deaths under status quo conditions and subject to interventions like lowering the prescription opioid supply.
\citet{pitt} and \citet{Rao2021} aim to project overdose deaths, life years, and quality-adjusted life years for several different policy responses (\ie, reducing opioid prescribing rates, expanding excess opioid disposal programs) using a compartmental model. The effect of a policy intervention is simulated by varying compartmental model parameters based on an ``assumed magnitude'' of impact and then projecting future outcomes~\citep{pitt, Rao2021}. In contrast, within our approach, we directly connect specific proposed policy decisions to exact changes in our compartmental model parameter values by integrating a dynamical model of the opioid epidemic into an optimization problem. To our knowledge, no previous work takes this approach within the context of the opioid epidemic. Rather than the traditional approach of assessing the effectiveness of a broad swath of policies, we provide a streamlined decision-making process for one type of intervention related to improving opioid treatment access.

For other applications, there have been previous efforts to integrate epidemiological models and optimization methods to inform policy decisions. \citet{Rao2021-yw} and \citet{10.1093/imammb/19.4.235} use compartmental models to inform simple optimization routines that can be solved using heuristics for vaccine and budget allocation, respectively. \citet{vax_location} integrate a compartmental model of the COVID-19 pandemic, called the DELPHI model, into a prescriptive optimization problem to make decisions about the optimal locations of mass vaccination facilities and optimally allocate COVID vaccines. They formulate a bilinear non-convex optimization problem which includes a modified version of the time-discretized DELPHI model within its constraints~\citep{vax_location}. Our work builds on the DELPHI model and this framework by specializing to the opioid epidemic context.
Additionally, for our compartmental model of the opioid epidemic, we take a different approach to parameter estimation by using \secondReviewChanges{a model-fitting process inspired by} neural ODEs. We also estimate unique model parameters for each state and formulate separate MIPs for each state, rather than having a single national-level optimization problem as in \citet{vax_location}. In doing this, our approach explicitly takes into account the unique opioid epidemic dynamics of each state, allowing for more targeted county-level solutions. Rather than only considering population-based equity, our MIPs also aim to ensure that the allocation of treatment facilities in a state is more socioeconomically equitable by considering the social vulnerability of each county.

\subsection{Our Contributions}
\begin{figure}[htb]
\centering
\ifpreprint
\includegraphics[width=.7\textwidth,trim={1.5cm 1cm 1.5cm 2cm},clip]{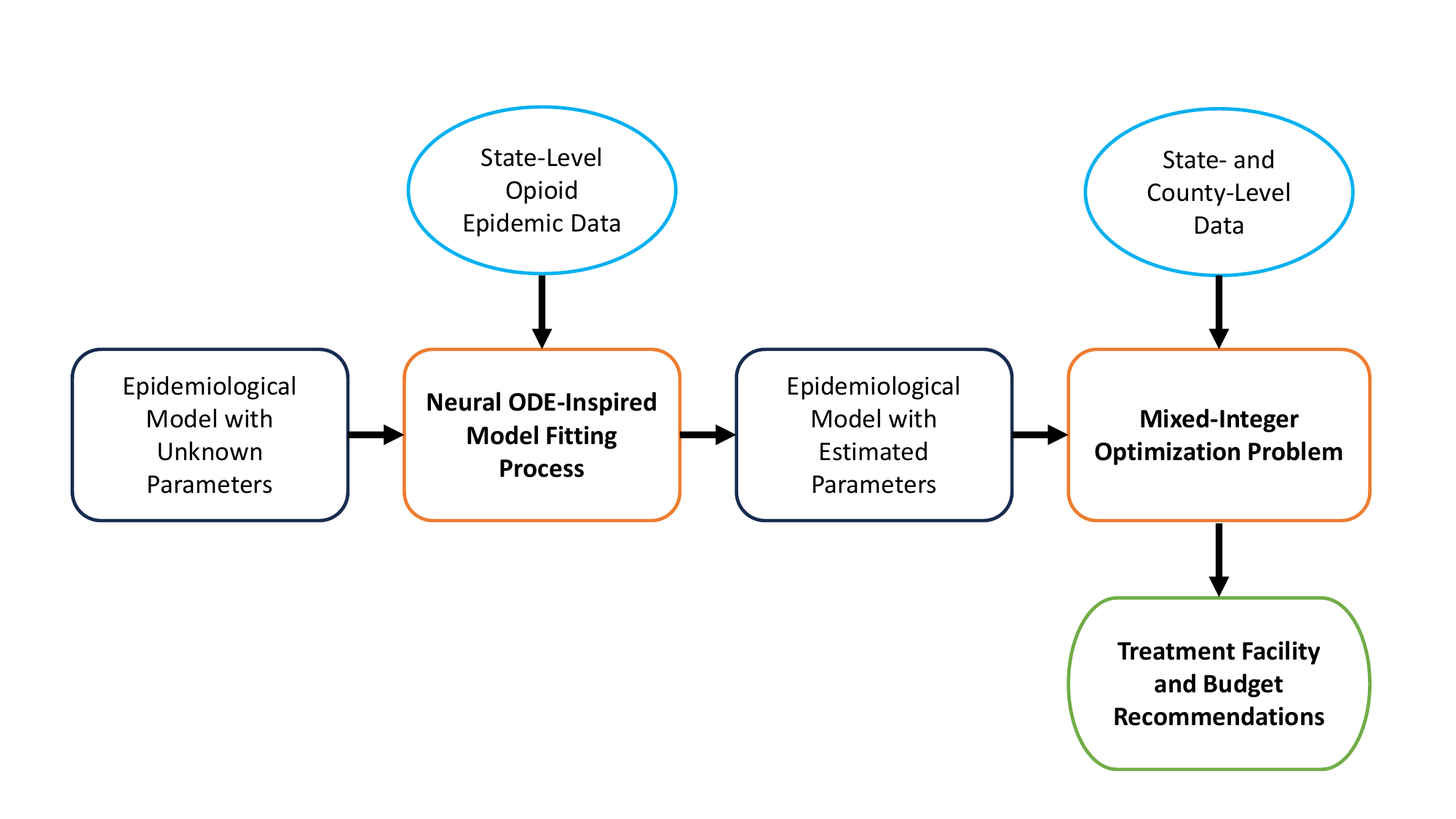}
\else
\includegraphics[width=.55\textwidth,trim={1.5cm 1cm 1.5cm 2cm},clip]{./Figures/model_process_flowchart.pdf}
\fi
\caption{Flow diagram summarizing our approach. Blue circles indicate data sources, orange blocks our Neural ODE and Mixed-Integer Optimization models, and the green block the resulting decisions.}
\label{fig:overall_flow}
\end{figure}

We present several methodological and operational contributions. Firstly, we seamlessly integrate a predictive dynamical model and a prescriptive optimization problem to create an operationally viable and streamlined approach for more equitable opioid treatment facility and treatment budget allocation. \secondReviewChanges{A flowchart summarizing the approach is shown in Figure~\ref{fig:overall_flow}.} \reviewChanges{We develop strong relaxations based on McCormick envelopes to efficiently obtain high-quality solutions with \secondReviewChanges{a mean optimality gap of 3.99\% (10.39\% maximum).}}
This approach is novel within opioid epidemic modeling and policy literature. Secondly, we show that a simple \secondReviewChanges{neural ODE-inspired} model informed by a dynamical structure can accurately estimate interpretable parameters from sparse time series data in the context of the opioid epidemic. These interpretable parameters can help quantify the differences between the opioid epidemic dynamics of different states. Finally, in terms of operational contributions, we show that optimizing opioid treatment facility and treatment budget allocation could have a positive impact, even in the short term, on population health measures like the number of people with OUD, the number of people receiving treatment, and the number of overdose deaths. \secondReviewChanges{Our optimized method also has a greater impact on these population health measures compared to population- and social vulnerability-based benchmarks.}
This work could help support future decision-making efforts related to improving opioid treatment access. Reproduceable code can be found \ifpreprint at: \url{https://github.com/joyceluo1/mip_opioid}. \else at: \url{https://github.com/joyceluo1/mip_opioid}. \fi

\section{Epidemiological Model}\label{epi}
\subsection{Model Definition} \label{def}
We formulate a general US state-level compartmental model, partitioning the population of a state into the following 6 exhaustive population classes (compartments):
\begin{itemize}
    \item \textbf{Susceptible (S):} Individuals who are not using opioids.
    \item \textbf{Prescribed (P):} Individuals who use or misuse prescription opioids but are not addicted.
    \item \textbf{Illicit Use (I):} Individuals who use illicit opioids like heroin.
    \item \textbf{Addicted (A):} Individuals who are addicted to prescription or illicit opioids.
    \item \textbf{Rehabilitating (R):} Individuals who are getting treatment for their addiction.
    \item \textbf{Deceased (D):} Individuals who have died from opioid overdoses.
\end{itemize}
\begin{figure}[htb]
\centering
\includegraphics[scale = .3, trim={2.5cm 1cm 4cm 1.2cm},clip]{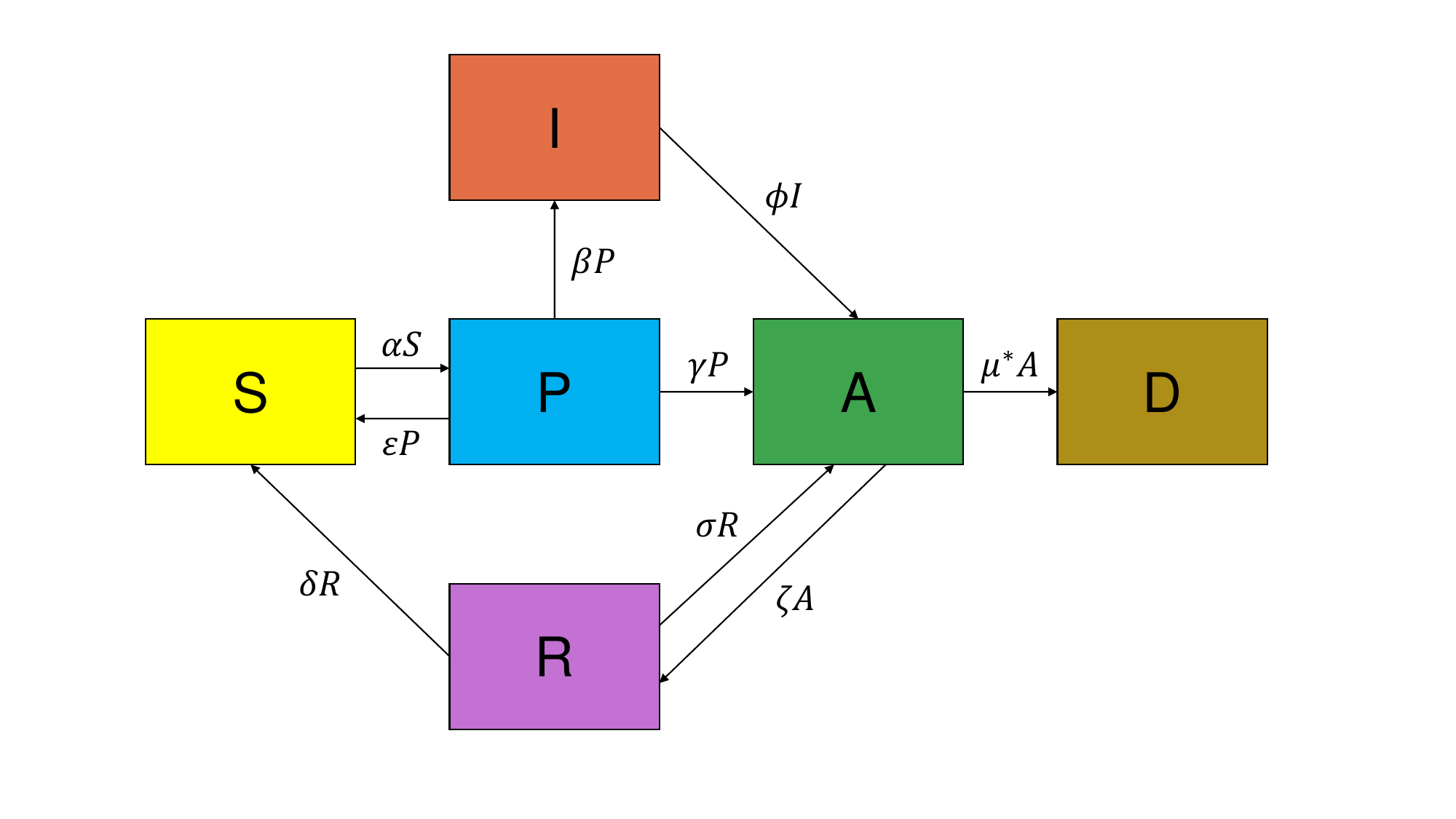}
\caption{Flow diagram of our state-level compartmental ODE model of the opioid epidemic.}
\label{fig:flow}
\end{figure}

Figure~\ref{fig:flow} shows a flow diagram of the compartmental model for a particular state. The model schematic shows the different population compartments, and the arrows depict how individuals transition between these different compartments. This deterministic model can be represented by a system of ODEs, which depends on 9 parameters. These parameters are the rates by which individuals move from one compartment to another in the model, as illustrated by their locations on particular arrows in Figure~\ref{fig:flow}. We assume that each state has unique opioid epidemic dynamics and therefore a unique model parameterization. We model the opioid epidemic using the system
$$\frac{\rm{d}\textbf{z}}{\rm{d} \textit{t}} = f(\textbf{z}(t),\rho, t),$$
with state vector $\textbf{z}(t) = (S(t), P(t), I(t), A(t), R(t), D(t)) \in \reals^6$, and we have the initial condition $\textbf{z}(0) = \textbf{z}_0 = (S_0, P_0, I_0, A_0, R_0, D_0)$ consisting of estimated data for each compartment from the year 1999 for a particular US state. Vector $\rho \in \reals^9$ represents the parameters that determine how the process evolves over time. The dynamics are represented by the following system of ODEs:
\begin{align*}
\frac{\rm d\textit{S}}{\rm d\textit{t}} & = \epsilon P + \delta R - \alpha S \\
\frac{\rm d\textit{P}}{\rm d\textit{t}} & = \alpha S - (\epsilon + \gamma + \beta)P \\
\frac{\rm d\textit{I}}{\rm d\textit{t}} & = \beta P - \phi I  \\
\frac{\rm d\textit{A}}{\rm d\textit{t}} & = \gamma P + \sigma R + \phi I - \zeta A - \mu A \\
\frac{\rm d\textit{R}}{\rm d\textit{t}} & = \zeta A - (\delta + \sigma)R \\
\frac{\rm d\textit{D}}{\rm d\textit{t}} & = \mu A,
\end{align*}
where $N = S + P + I + A + R + D$ is the total population of a particular US state. We assume that interactions between compartments are linear, as non-linear interactions were deemed negligible through parameter estimation.
We estimate the unknown parameters in $\rho$ from real world data to ensure that our dynamics function $f(\textbf{z}(t),\rho, t)$ approximates the true dynamics closely.

\subsection{Parameters}
The epidemic model is based on parameters $\rho = (\alpha, \gamma, \delta, \sigma, \mu, \xi, \epsilon, \phi, \beta)$ described in Table~\ref{tab:model_params}.
\begin{table}[htb]
\ifpreprint \else
    \TABLEsmallX
    \fi
    \begin{minipage}{\textwidth}
    \centering
    \caption{Model parameters $\rho = (\alpha, \gamma, \delta, \sigma, \mu, \xi, \epsilon, \phi, \beta) \in \reals^9$.}
\begin{tabular}{@{}ll@{}}
\toprule
 & Description\\
\midrule
$\alpha$ & Prescription rate per person per year\\
$\gamma$ & Prescription-induced addiction rate\\
$\delta$ & Successful treatment rate\\
$\sigma$ & Natural relapse rate of an individual receiving treatment\\
$\mu$  & Death rate of addicts\\
$\zeta$  & Rate of individuals with OUD entering into rehabilitation\\
$\epsilon$ & Rate of ending prescription without addiction\\
$\phi$ & Illicit drug-induced addiction rate\\
$\beta$ & Transition rate from prescription to illicit opioid use\\
\bottomrule
\end{tabular}
    \label{tab:model_params}
    \end{minipage}
\end{table}
All parameters represent constant annual transition rates between particular compartments.
We assume the parameters are time invariant~\citep{battista}. These parameters are similar to those within previous literature~\citep{battista, pitt}, but we add parameters which take into account the effect of illicit opioids on the dynamics of the opioid epidemic. In particular, we are interested in the illicit drug-induced addiction rate ($\phi$) and how people transition from prescription to illicit opioid use ($\beta$). In a previous iteration of the model, we also considered the illicit opioid use initiation rate. However, through parameter estimation, we determined that this parameter had a negligible effect on the dynamics of the model, and it was removed. Therefore, we assume that individuals can only initiate illicit use if they previously used prescription opioids. This is also substantiated by previous research indicating that the majority of heroin users have misused prescription opioids in the past~\citep{muhuri,LANKENAU201237}.
We set parameters $(\alpha, \gamma, \delta, \sigma) = (0.15, 0.00744, 0.1, 0.9)$ based on \citet{battista} and estimate parameters $(\phi, \epsilon, \beta, \zeta, \mu)$ using our \secondReviewChanges{neural ODE-inspired model fitting process}.


\subsection{Data}
We collected state-level data for the years 1999 to 2019 for compartments $D$, $P$, $I$, $A$, and $R$ to estimate the unknown model parameters. We define the time horizon to be $H=20$ (\ie, the year 1999 represents $t=\tau = 0$ and the year 2019 represents $\tau = H$, where $\tau$ is a discrete time unit from $0,1,\dots,H$). The Multiple Cause-of-Death dataset in the CDC WONDER (Wide-ranging Online Data for Epidemiologic Research) database~\citep{cdcwonder} was our source for yearly overdose death counts ($D$) from 1999--2019. To identify opioid-specific deaths, we filtered the dataset using the multiple cause-of-death (ICD) codes for heroin (T40.1), natural opioid analgesics (T40.2), methadone (T40.3), and synthetic opioid analgesics other than methadone (T40.4). We also used underlying cause-of-death codes X40–X44 (unintentional), X60–X64 (suicide), and Y10–Y14 (undetermined). The CDC suppresses data values below a threshold of 10 to prevent patient identification, so we removed states if over 20\% of their data was suppressed. The following states were removed: North Dakota, South Dakota, Alaska, Idaho, Montana, Mississippi, Wyoming, and West Virginia. For states with fewer suppressed measures, these were replaced with an integer between 1 and 9 drawn from a uniform distribution. We made the deaths cumulative, starting from the number of deaths in 1999. This is because we assume that the Deceased compartment is absorbing, since those who die from opioid overdoses cannot transition into other compartments.

We approximated the number of people using prescription opioids per year per state ($P$) using data sources from the CDC. The CDC provides data regarding opioid dispensing rates for each state from 2006 to 2019~\citep{dispense_rates}. We used this data to calculate ratios of the number of prescription opioids dispensed in each state to the number dispensed nationally. In addition, the CDC's National Health and Nutrition Examination Survey (NHANES) provides biyearly estimates of the percentage of adults nationwide who used a prescription opioid in the past 30 days for the years 1999--2018~\ifpreprint \citep{nhanes}. \else \citepalias{nhanes}. \fi From these percentages, we estimated the \textit{number} of adults nationwide who used a prescription opioid per year. To obtain state-level estimates, we multiplied our state-to-national ratios by the nationwide estimate calculated from NHANES for each year from 2006--2018.

From SAMHSA's National Survey on Drug Use and Health (NSDUH), we obtained data on the yearly prevalence of OUD ($A$) for each state from 2016--2019. We also obtained the yearly estimated number of heroin users for each state from 2016--2019 ($I$)~\ifpreprint \citep{NSDUH}. \else \citepalias{NSDUH}. \fi For the number of people in treatment ($R$), we obtained data from the National Survey of Substance Abuse Treatment Services for the years 2000, 2002--2013, 2015--2017, and 2019~\ifpreprint \citep{NSSATS}. \else \citepalias{NSSATS}. \fi The data measure was the aggregated number of clients receiving MAT across all facilities in a state within a day each year.

We calculated the number of susceptible people ($S$) based on data from the other 5 compartments and the populations of each US state. We assume that $N^{(\tau)} = S^{(\tau)} + P^{(\tau)} + I^{(\tau)} + A^{(\tau)} + R^{(\tau)} + D^{(\tau)}$, where $N^{(\tau)}$ is the state population in year $\tau$. The death counts are included in this summation because they are negligible compared to the total population. We implicitly consider overall birth and death rates of the population by allowing $N^{(\tau)}$ to vary. We calculated the $S$ compartment only for the time points with complete data for all other compartments. For other time points, the $S$ value was set to 0. We detail how we set the model initial conditions in \ifpreprint Appendix~\ref{ic}. \else Section EC.1. \fi

We created data matrices for every included US state. Each data matrix is $21 \times 6$, and each row represents $\textbf{z}^{(\tau)} \in \reals^6$, the data observation at time $t = \tau$ for $\tau = 0,1,\dots,H$.
Missing compartment values were set to 0. In order to ensure convergence and a better model fit, we normalized each compartment's value by $N^{(\tau)}$ at each time point.

\subsection{Neural ODE-Inspired Model Fitting Process}\label{node}


\secondReviewChanges{The neural ODE framework represents ODEs and their solvers as a neural network layer in combination with more traditional neural network layers \citep{chen2019neural}}.
ODEs and ODE solvers fit perfectly into the neural network framework, as they have been proven to be differentiable~\citep{chen2019neural}.
\secondReviewChanges{We apply a simplified version of the neural ODE framework, where we use the structured ODE model defined in Section~\ref{def} as the single layer in our neural network. We then estimate the unknown parameters of our ODE-based model $(\phi, \epsilon, \beta, \zeta, \mu)$ by training this simple neural network.}
\secondReviewChanges{Although this model can be expanded to include more neural network layers, we only use the structured ODE dynamics to ensure interpretability of the model and its estimated parameters.} We use gradient-based optimizers to minimize the following \secondReviewChanges{average-weighted 2-norm loss}:
\secondReviewChanges{$$\loss((\phi, \epsilon, \beta, \zeta, \mu)) = \sum_{\tau=0}^{H}
\| q_{\tau} \odot (\hat{\textbf{z}}^{(\tau)} - \textbf{z}^{(\tau)})\|_2^2,$$}where $\hat{\textbf{z}}^{(\tau)}$ is the ODE model's prediction of the vector of compartment values and $\textbf{z}^{(\tau)}$ is our observation of the vector of compartment values from the data at time $t=\tau$ for $\tau = 0,1,\dots,H$. The vectors $\hat{\textbf{z}}^{(\tau)}$ and $\textbf{z}^{(\tau)}$ lie in $\reals^6$. 
We define $q_{\tau} \in \reals^6$ to penalize differences between the predictions and observations based on data availability \secondReviewChanges{and the magnitudes of the differences} at time point $\tau$. If there is no data for a compartment at $\tau$, the corresponding element in $q_{\tau}$ is set to 0. 
However, if the rest of the compartment data at $\tau$ is available, the corresponding elements in $q_{\tau}$ are set to \secondReviewChanges{the reciprocal of the average size across time for each compartment}, and only those elements are used to calculate the loss. 
Here, $\odot$ denotes the element-wise product of $q_{\tau}$ and $\hat{\textbf{z}}^{(\tau)} - \textbf{z}^{(\tau)}$.

In contrast to traditional neural networks, \secondReviewChanges{this model fitting process inspired by neural ODEs} does not require a large amount of data to estimate parameters accurately, which makes it ideal for applications with limited data like is the case for the opioid epidemic. It also allows us to estimate parameters in a more computationally efficient way, because we are able to estimate the parameters based on the direction of the gradient~\citep{chen2019neural}.

\paragraph{Implementation.} We use Julia to implement our \secondReviewChanges{neural ODE-inspired model fitting process}. In particular, we use the \texttt{DiffEqFlux.jl}~\citep{DBLP:journals/corr/abs-1902-02376} and \texttt{DifferentialEquations.jl}~\citep{DifferentialEquations.jl-2017} libraries. We train our model using this method for each individual state.
Our initial condition is a vector of the normalized compartment values in 1999 for each respective state.
We set an initial guess for the unknown parameters: ${(\phi, \epsilon, \beta, \zeta, \mu) = (0.3, 0.9, 0.19, 0.5, 0.01159)}$, according to parameter ranges and estimates from previous literature~\citep{battista}.
We restrict the parameters to be non-negative by representing them as the square of the actual parameters we estimated (\eg, $\phi = \hat{\phi}^2$, where $\hat{\phi}$ is the actual model parameter that we learn). We use the ODE solver \texttt{Tsit5} and the \texttt{ForwardDiffSensitivity} method to calculate the gradients. To perform stochastic gradient descent, we run the ADAM optimization algorithm for 25000 iterations with a step size of 0.0001, followed by BFGS.


\section{Mixed-Integer Optimization Problem}\label{mip}
The overarching goal of our approach is to offer solutions to ensure that MAT and treatment facilities are more accessible and allocated equitably. Accordingly, for each US state, we formulate a prescriptive MIP to address two main objectives: opioid treatment facility location and treatment budget allocation. In particular, we focus on treatment facilities that offer MAT. Our MIPs mainly aim to minimize overdose deaths and the number of people with OUD, but also take into account socioeconomic considerations so that treatment facilities are distributed more equitably. The initial starting point for our problem is 2017, as that year has sufficient data availability. We set our modeling period to be 2 years.

\subsection{Data}\label{mip_data}
We obtained data related to the current number of treatment facilities that offer MAT in each county, using the SAMHSA Behavioral Health Treatment Services Locator. This tool helped us create a dataset indicating the number of treatment facilities that offered ``Outpatient methadone/buprenorphine or naltrexone treatment'' in each county~\citep{b_locator}. We needed this data to account for the effect of facilities that are already treating patients with MAT. We also obtained data from the CDC's Social Vulnerability Index (SVI), which provides a value for each county that captures 15 factors from the US Census, including poverty, lack of vehicle access, and crowded housing~\ifpreprint \citep{SVI}. \else \citepalias{SVI}. \fi This index is intended to help identify populations that are vulnerable during public health emergencies like the opioid epidemic. The SVI ranking is a value between 0 and 1, with a ranking closer to 1 indicating that the region is more socially vulnerable~\ifpreprint \citep{SVI}. \else \citepalias{SVI}. \fi The CDC provides SVI data every 2 years, and we obtained county-level data for 2018. Additionally, we obtained county-level data regarding opioid dispensing/prescribing rates per 100 people for 2018~\citep{dispense_rates} and county population totals from the Census Bureau~\citep{county_pop}.

Budgetary information for each state was obtained for the constraints of our MIPs. We obtained total grant funding data for each state in 2018 from the US Department of Health and Human Services (HHS) Opioid Grants Dashboard~\citep{total_bud}. According to previous opioid grant spending analyses, around 65\% of grant funding was used for treatment initiatives in a particular year~\citep{samhsa_grant}. In addition, the estimated cost of opening a treatment facility ranges from \$300--600K for an intensive outpatient facility~\citep{otp_cost}. We rounded up the cost to \$1,000,000 to have a higher estimate. For each state, we budgeted 65\% of the total grant funding to be used for opening new treatment facilities and divided this number by \$1,000,000 to get the maximum \textit{additional} number of treatment facilities that can be opened in that state. We added this number to the current number of treatment facilities to get cap on the number of facilities that can be opened in that state, which we call $N$.

SAMHSA recently distributed a grant to states for the purposes of expanding MAT~\citep{samhsa_grant}. We divided the amount distributed to each state by 4 to get quarterly estimates of the treatment budget, which we call $d_k$ at time $k$. For the scope of this work, we chose to only focus on methadone-based MAT. Accordingly, we also obtained data on the weekly cost of treating a patient with methadone-based MAT, which was \$37.38~\citep{med_cost}. Multiplying this number by 12 gave us $d = \$448.56$, the quarterly cost of methadone-based MAT. In future iterations of this MIP, we could additionally take into account buprenorphine-based MAT.

\subsection{Problem Formulation}\label{prob}
\secondReviewChanges{\paragraph{Assumptions.} Over our modeling period of 2 years, we assume that the opening of facilities is a static decision made at the beginning of the time horizon but the allocation of the treatment budget happens dynamically over time. We also assume that our decisions can concretely affect our estimated epidemiological model parameters (\ie, adding treatment capacity is highly correlated with more patients getting treated). Additionally, we assume that patients would rather stay within their own county to get treatment, so there should be at least one facility in each county. } \thirdReviewChanges{However, we relax this last assumption in an alternative formulation that incorporates spatial information; see \ifpreprint Appendix~\ref{spatial}. \else Section EC.7. \fi}
\paragraph{Parameters and Decision Variables.} We consider an optimization problem over a time horizon $K$ with time periods $k \in \mathcal{K} = \{0,\dots,K\}$.
We denote each county in a state as $i$ with $i \in \mathcal{C} = \{1,\dots,C\}$, where $C$ is the total number of counties in a state.
The decision variables are $x_i$---denoting the number of opioid treatment facilities with MAT needed for county $i\in\mathcal{C}$, and $\bar{d}_{ik}$---denoting the treatment budget distributed to county $i\in\mathcal{C}$ at time $k \in \mathcal{K}$.
The optimization problem will feature the parameters in Table~\ref{tab:prob_params}.
\begin{table}[htb]
\ifpreprint \else
    \TABLEsmallX
    \fi
    \begin{minipage}{\textwidth}
    \centering
    \caption{Parameters of the optimization problem.}
    \begin{tabular}{@{}ll@{}}
    \toprule
     & Description\\
    \midrule
    $n_i$ & Number of opioid treatment facilities with MAT already in county $i \in \mathcal{C}$ \\
    $\mathrm{SVI}_i$ & Social Vulnerability Index in county $i \in \mathcal{C}$ \\
    $\mathrm{pr}_i$ & Prescribing rate per 100 people in county $i \in \mathcal{C}$ \\
    $\mathrm{Pop}_i$ & Population in county $i \in \mathcal{C}$ \\
    $N$ & Maximum number of treatment facilities that can be open in the state \\
    $d_k$ & Treatment budget limit for time $k \in \mathcal{K}$ \\
    $d$ & Quarterly per patient cost of MAT \\
    $\alpha$ & Prescription rate per person per year \\
    $\gamma$ & Prescription-induced addiction rate \\
    $\delta$ & Successful treatment rate \\
    $\sigma$ & Natural relapse rate of an individual in treatment \\
    $\mu$ & Death rate of addicts \\
    $\zeta$ & Rate of individuals with OUD entering into rehabilitation \\
    $\epsilon$ & Rate of ending prescription without addiction \\
    $\phi$ & Illicit drug-induced addiction rate \\
    $\beta$ & Transition rate from prescription to illicit opioid use \\
    \bottomrule
    \end{tabular}
    \label{tab:prob_params}
    \end{minipage}
\end{table}

To use the continuous time compartmental model described in Section~\ref{def} within the constraints of our optimization problem, we discretize it using the forward Euler method~\citep{discre}. We set $t = k\Delta$, where $\Delta$ is the time discretization interval and $k\in\mathcal{K}$.
In order to mimic the continuous time trajectory as closely as possible while also not making $\Delta$ too small, we set $\Delta = 0.25$. This represents a time increment of 3 months.

Our decision variables regarding opening additional treatment facilities and establishing treatment budgets for counties act as proposed interventions that affect state-level opioid epidemic dynamics (\ie, the compartmental model parameters).
We model the impact of our decision variables on state-level opioid epidemic dynamics by showing how they affect the estimated parameter $\zeta$ in each time period.
Having greater access to opioid treatment facilities that offer MAT helps more people who have OUD get the treatment they need. Therefore, optimizing the treatment facility and treatment budget allocation should increase $\zeta$, the transition rate from the $A$ to $R$ compartment. We affect $\zeta$ in each time period based on $\ell_k$, the proportion of extra people that could transition from $A$ to $R$ if a certain number of new treatment
facilities offering MAT were established. We define this proportion as
$$\ell_k = \frac{\sum_{i = 1}^C (x_i - n_i)\bar{d}_{ik}/dx_i}{A_k}.$$ 
\secondReviewChanges{We have $\bar{d}_{ik}/x_i$ representing the treatment budget distributed to each facility in a county, which is dependent on two of the decision variables. We ensure that the treatment budget per facility must be greater than a certain minimum budget to reduce the effect of diminishing returns as the number of facilities $x_i$ increases. We show the details of this when describing our constraints. Dividing this quantity by $d$, the cost to fully treat a patient with MAT for a quarter of the year, gives $\bar{d}_{ik}/dx_i$ -- the number of patients treated per facility per quarter within a county. Multiplying this by the additional number of facilities in a county gives the additional number of people who could be treated due to the new treatment facilities. We then sum this quantity over all counties and divide by $A_k$ to get the added rate of transition from the $A$ to $R$ compartments for the state.}
For a particular state, we have the following dynamics:
\begin{equation*}
\begin{array}{ll}
    S_{k+1} &= S_k + (\epsilon P_k + \delta R_k - \alpha S_k)\Delta\\
    P_{k+1} &= P_k + (\alpha S_k - (\epsilon + \gamma + \beta)P_k)\Delta \\
    I_{k+1} &= I_k + (\beta P_k - \phi I_k)\Delta\\
    A_{k+1} &= A_k + (\gamma P_k + \sigma R_k + \phi I_k - \left(\zeta + \ell_k\right) A_k - \mu A_k)\Delta \\
    R_{k+1} &= R_k + \left(\left(\zeta + \ell_k\right) A_k - (\delta + \sigma)R_k\right)\Delta\\
    D_{k+1} &= D_k + (\mu A_k)\Delta.
        \end{array}
\end{equation*}
Collectively, we define this approximated compartmental dynamics function as $\bar{f}$.
For a particular US state, we define the following treatment facility location and budget allocation problem:

\secondReviewChanges{
\begin{equation*}\label{eq:mip_implicit}
\begin{array}{ll}
\ifpreprint
        \text{minimize} & \lambda_D D_K + \lambda_A\sum_{k = 0}^K (A_k) + \lambda_{\mathrm{pr}}\mathbf{c}_{\mathrm{pr}}^{\top}\mathbf{x} + \lambda_{\mathrm{SVI}}\mathbf{c}_{\mathrm{SVI}}^{\top}\mathbf{x}  + \lambda_{\mathrm{Pop}}\sum_{k = 0}^K \mathbf{c}_{\mathrm{Pop}}^{\top}\mathbf{\bar{d}}_{k} \\
      & \textstyle\hspace{19.8em}+ \lambda_{\mathrm{inf}}\max\left(0, \sum_{i = 1}^C x_i - N\right) \\
          \else
          \text{minimize} & \lambda_D D_K + \lambda_A\sum_{k = 0}^K (A_k) + \lambda_{\mathrm{pr}}\mathbf{c}_{\mathrm{pr}}^{\top}\mathbf{x} + \lambda_{\mathrm{SVI}}\mathbf{c}_{\mathrm{SVI}}^{\top}\mathbf{x} + \lambda_{\mathrm{Pop}}\sum_{k = 0}^K \mathbf{c}_{\mathrm{Pop}}^{\top}\mathbf{\bar{d}}_{k} + \lambda_{\mathrm{inf}}\max\left(0, \sum_{i = 1}^C x_i - N\right) \\
          \fi
	    \text{subject to} &  x_i \geq n_i, \quad  \forall i \in \mathcal{C}\\
		& \sum_{i=1}^C \bar{d}_{ik} \leq d_k, \quad   \forall k \in \mathcal{K}\\
        & \bar{d}_{ik}/x_i \geq d_{\mathrm{min}}, \quad
            \forall i \in \mathcal{C}, k \in \mathcal{K} \\
         & \bar{d}_{i,k-1} - \bar{d}_{ik} \leq \theta , \quad
    \forall i \in \mathcal{C}, k \in 1, \dots, K \\
  	& \mathrm{lb}_{ik} \leq \bar{d}_{ik} \leq \mathrm{ub}_{ik}, \quad
        \forall i \in \mathcal{C},k \in \mathcal{K}\\
		& {\rm\bf z}_{k+1} = \bar{f}({\rm\bf z}_k, {\rm\bf x}, \bar{{\rm\bf d}}_k), \quad k=0,\dots,K-1\\
        &x_i \in \integers, \quad x_i \geq 1, \quad   \forall i \in \mathcal{C}\\
        & {\rm\bf z}_k = (S_{k}, P_{k}, I_{k}, A_{k}, R_{k}, D_{k}) \ge 0,\quad \forall k \in \mathcal{K}\\
        &\bar{d}_{ik} \ge 0, \quad \forall i \in \mathcal{C},k \in \mathcal{K}.
        \end{array}
\end{equation*}
}

\paragraph{Objective.}
Within the MIP objective function, we prioritize minimizing the total number of overdose deaths and the total number of people with OUD, which are the first two terms of the objective.
\reviewChanges{In the next two terms, we define costs $\mathbf{c}_{\mathrm{SVI}}$ and $\mathbf{c}_{\mathrm{pr}}$ to ensure that $\mathbf{x}$ (the vector of decision variables for the number of treatment facilities per county) is more proportional to the distributions of SVI rankings and prescribing rates across counties. We define the cost for a county as the inverse of its SVI ranking. This makes the cost high when the SVI ranking is low and vice versa. Therefore, this ensures that more facilities will be allocated to more vulnerable counties (high SVI ranking) in order to minimize the costs resulting from this term. For the prescribing rate costs, we use min-max normalization on the prescribing rates to ensure all the values are between 0 and 1, and then take the inverse of these normalized values. This similarly ensures that more facilities are allocated to counties with higher prescribing rates. The parameters $\lambda_{\mathrm{pr}}$ and $\lambda_{\mathrm{SVI}}$ determine whether the allocation of treatment facilities is more proportional to the prescribing rate distribution or the SVI ranking distribution. These terms ensure that the solutions take into account socioeconomic equitability. \secondReviewChanges{We also take this approach for the treatment budget in each time period, defining $\mathbf{c}_{\mathrm{Pop}}$ as the inverse of the county population proportion.} The final term of the objective function is essential for solution feasibility. In most cases, $\sum_{i = 1}^C x_i = N$, as we aim to cap the total number of treatment facilities in the state at $N$. If $\sum_{i = 1}^C x_i > N$, this indicates that the state's solution must be over budget to be feasible. In order to minimize the amount that a solution is over budget, we penalize $\sum_{i = 1}^C x_i - N$ by setting a larger $\lambda_{\mathrm{inf}}$. This objective function approach gives us more insight into potential solutions if particular states exceed their budget.}

\paragraph{Constraints.}
The first constraint ensures that the recommended number of treatment facilities in a county is greater than or equal to the number of treatment facilities already in that county. The second constraint limits the sum of the treatment budgets distributed to each county by the state treatment budget limit for time $k$. \secondReviewChanges{The third constraint ensures that the budget per facility within county $i$ at time $k$ is greater than a minimum budget $d_{\mathrm{min}}$, which we set to \$5,000 to ensure that at least 10 patients per facility can be fully treated per quarter. We use that the cost of treating a patient per quarter is $\sim\$500$ from a~\citet{med_cost} fact sheet. The fourth constraint ensures that the treatment budget can only decrease by at most $\theta$ between time periods. We set $\theta = \$10,000$ to ensure the budget cannot decrease drastically. The fifth constraint sets the bounds on the treatment budget to ensure that counties are getting at least a certain budget per quarter. We define $\mathrm{lb}_{ik}$ and $\mathrm{ub}_{ik}$ as the following:
$$\mathrm{lb}_{ik} = d_{\mathrm{min}} n_i \text{ if } n_i \geq 1, \quad d_{\mathrm{min}} \text{ otherwise}; \quad \mathrm{ub}_{ik} = d_{\mathrm{max}}(n_i + a). $$
The cases for ${lb}_{ik}$ ensure resources are allocated even when $n_i=0$, because since $x_i \geq 1$, at least one facility will be allocated to county $i$ in which the current number of facilities is 0. The amount $a$ represents the total additional number of facilities we can allocate according to the state budget. We set $d_{\mathrm{max}}$ to a large enough value to ensure that the optimization solution can use the entire treatment budget limit $d_k$ for that period. We describe this selection process further in \ifpreprint Appendix~\ref{state_reform}. \else Section EC.2. \fi}
The sixth constraint describes the discretized compartmental model for the particular state. The seventh constraint restricts the $x_i$'s to be integer-valued and requires that there is at least one treatment facility in each county. This ensures that there will not be any issues with dividing by 0, as we divide by $x_i$ within our $\ell_k$ term. The eighth and ninth constraints describe the domain of the compartment values at each time $k$, and the domain of the $\bar{d}_{ik}$ decision variables. The problem as shown here is not easily solved by a numerical MIP solver, since we are dividing two decision variables and have a non-convex objective. We reformulate the problem to have at most quadratic constraints and a linear objective, and we scale the decision variables (shown in \ifpreprint Appendix~\ref{state_reform}). \else Section EC.2). \fi 

\thirdReviewChanges{
\paragraph{Incorporating Spatial Information.}
Our formulation can be extended to take into account spatial information. Since patients could potentially use resources from neighboring counties, we develop an alternative formulation which considers that certain counties do not need to have treatment facilities if a sufficient number of surrounding counties have facilities; 
see \ifpreprint Appendix~\ref{spatial}. \else Section EC.7. \fi

}

\reviewChanges{\subsection{Solution Method}\label{sol_method}
Unfortunately, the non-convex bilinear MIP described in Section~\ref{prob} can be solved only for small instances.
To overcome this limitation, we introduce strong relaxations based on McCormick envelopes~\citep{mccormick} to obtain high-quality solutions with small optimality gaps for all problem instances. \secondReviewChanges{Our optimality gaps are obtained by finding tight upper and lower bounds on the optimal objective values and calculating the percent differences between the bounds.}
We use McCormick envelopes to linearize and relax the bilinear terms in our original formulation; see \ifpreprint Appendix~\ref{mcc}. \else Section EC.3. \fi We then solve the relaxed problem to obtain a lower bound on the optimal objective value. In addition, the optimal $x_i$ and $\bar{d}_{ik}$ (facility and budget allocation solutions, respectively) from the relaxation will still satisfy all constraints from the bilinear problem apart from the compartmental model constraints. This is because the bilinear terms only appear as a result of the compartmental model constraints.
Since the compartmental model constraints are not restrictive, we can use the facility and budget allocation from the relaxation to propagate the opioid epidemic dynamics described by the original non-relaxed constraints and compute the compartment values over time. Then, using the computed compartment values and the now-feasible facility and budget allocation, we can calculate what the objective value with the feasible solution would be to obtain an upper bound.
While we can obtain successively tighter upper and lower bounds using piecewise McCormick envelopes~\citep{KARUPPIAH2006650}, this is not necessary in our procedure because our \secondReviewChanges{mean optimality gap is 3.99\% (10.39\% maximum).} \thirdReviewChanges{Our method also obtains solutions much faster, taking 2.179 seconds on average compared to 521.220 seconds for the bilinear formulation.}
Detailed results are shown in \ifpreprint Appendix~\ref{opt_gap}. \else Section EC.10. \fi

\paragraph{Implementation.} We use Gurobi 10.0 \citep{gurobi} 
to solve MIPs for the bilinear formulation and our relaxed formulation for each state.
\secondReviewChanges{We set the following hyperparameters: $\lambda_D = 10$, $\lambda_A = 9$, $\lambda_{\mathrm{pr}} = 0.001$, $\lambda_{\mathrm{SVI}} = 0.009$, and $\lambda_{\mathrm{Pop}} = 0.001$. We set $\lambda_{\mathrm{inf}} = 10$ if $D_K < 10000$, and $\lambda_{\mathrm{inf}} = 9$, otherwise. }
Reproduceable code can be found \ifpreprint at: \url{https://github.com/joyceluo1/mip_opioid}. \else at: \url{https://github.com/joyceluo1/mip_opioid}. \fi
}

\section{Results and Discussion}\label{results}
\subsection{Epidemiological Model Parameter Estimation}
\begin{figure}[htb]
\centering
\includegraphics[width=.6\textwidth,trim={0 0 1cm 1cm},clip]{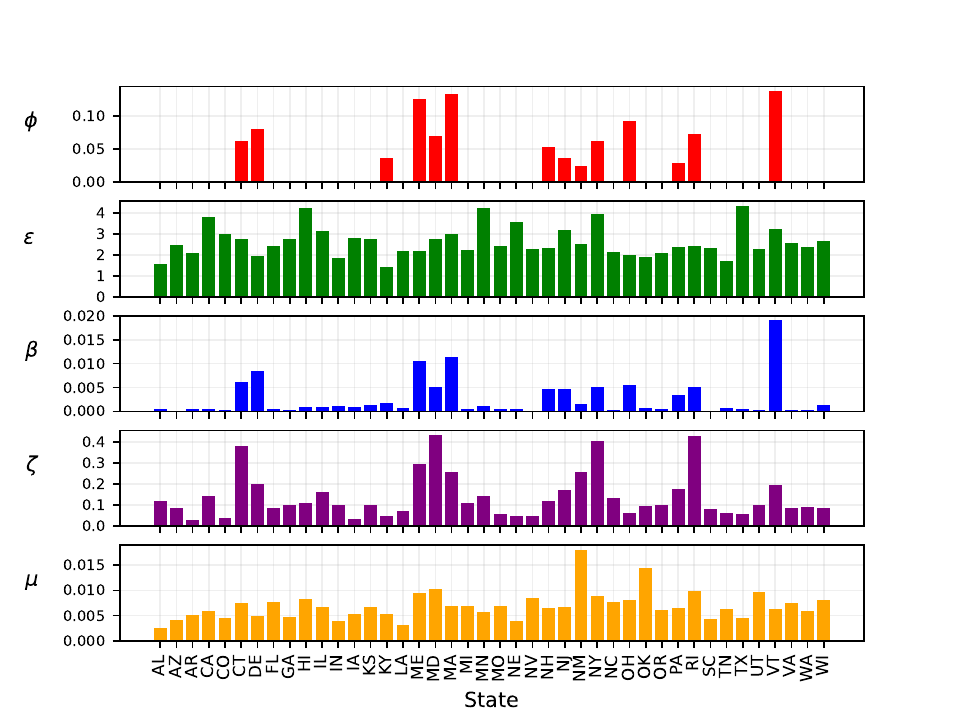}
\caption{Estimated parameters for each US state.}
\label{fig:params_fig}
\end{figure}

\secondReviewChanges{Figure~\ref{fig:params_fig} displays estimated parameters from our neural ODE-inspired model fitting process for each US state, with exact values included in \ifpreprint Appendix~\ref{tab_fig}. \else Section EC.11. \fi From Figure~\ref{fig:params_fig}, $\phi$ and $\beta$ show correlation, likely due to their shared relation to illicit opioid use. $\phi$ tends to be approximately 0 for many states, and $\beta$ has very small values, which is likely due to underestimation of illicit opioid use in NDSUH surveys. However, certain states have non-neglible $\phi$ and $\beta$ values. For New York, we see that 
\secondReviewChanges{$\phi = 0.0628$}, indicating a 6\% OUD risk for illicit opioid users. This is higher than the prescription-induced addiction rate $\gamma = 0.00744$, which we set based on previous literature~\citep{battista}. This makes sense because illicit opioids like heroin tend to be more addictive than prescription opioids. We see that $\beta$ = 0.0051 for New York, which suggests around 5 in 1000 New Yorkers who are using prescription opioids will begin using illicit opioids within a year. This seems to align with estimates which state that around 4--6\% of individuals who misuse prescription opioids transition to heroin~\citep{muhuri}. Since our $P$ compartment also includes people who properly use prescription opioids, it makes sense that our parameter estimate would be smaller. Vermont has the largest $\phi$ and $\beta$ values, possibly due to better data quality or an actual issue with people transitioning from prescription to illicit opioid use and becoming addicted. More states have non-negligible $\beta$ parameters compared to $\phi$, indicating the transition from prescription to illicit use contributes to the opioid epidemic. The parameters $\mu$, $\zeta$, and $\epsilon$ remain consistently non-negligible across states. The $\mu$ values align with previous death rate estimates \citep{battista} and are highest in Oklahoma (0.0145) and New Mexico (0.0181). The $\zeta$ values range from 0.03–-0.43, with Maryland having the largest. A~$\zeta$ value of 0.2 means that 20 addicted people enter treatment out of 100 addicted people. The $\epsilon$ values range from 1–-4 for different states, indicating most patients end their prescriptions without addiction within a year. Hawaii and Texas have the largest $\epsilon$ values. Our estimated parameters generally lie within ranges from the literature \citep{battista}.}

\secondReviewChanges{We also assess our model's performance using traditional metrics like mean absolute percentage error (MAPE) for each compartment and compare our model's performance with estimated parameters to its performance with parameters from the literature. Overall, the average MAPE across states for each compartment is low when we use our estimated parameters, and the MAPEs are considerably lower than when using the literature parameters.} Detailed numerical validation of the epidemiological model is shown in \ifpreprint Appendix~\ref{epi_val}. \else Section EC.4. \fi

\subsection{Optimization Problem Solutions}\label{sols}
Using our solution method, we obtain high-quality treatment facility and budget allocation solutions for the majority of the US states.
Figure~\ref{fig:sols} shows the additional treatment facilities and treatment budget allocations determined by the MIP solutions for California and Florida. 
Figures~\ref{fig:ca_x} and~\ref{fig:fl_x} show that most counties in each state either do not need any additional treatment facilities or only need one additional facility to be opened. 
\secondReviewChanges{
In the California MIP solution (Figure~\ref{fig:ca_x}), the most additional treatment facilities are recommended for counties like San Diego, Riverside, Los Angeles, and San Bernardino. In Florida's solution, the counties with the greatest recommended number of additional treatment facilities are Broward, Hillsborough, Miami-Dade, and Orange County. For both states, these indicated counties all have relatively large populations and SVI rankings of over 0.5. This means that making the facility allocations closer to the population share and high SVI rankings are both drivers of facility location patterns. High SVI rankings are a driver because we set the $\lambda_{\mathrm{SVI}}$ hyperparameter to be larger than $\lambda_{\mathrm{pr}}$ in the objective.
Further customization of facility location patterns can be achieved by adjusting these hyperparameters. 
}

\ifpreprint
\begin{figure}[htbp]
     \centering

     \begin{subfigure}[b]{0.49\textwidth}
        \centering
         \includegraphics[width=\textwidth, trim={6cm 1cm 0.5cm 1cm},clip]{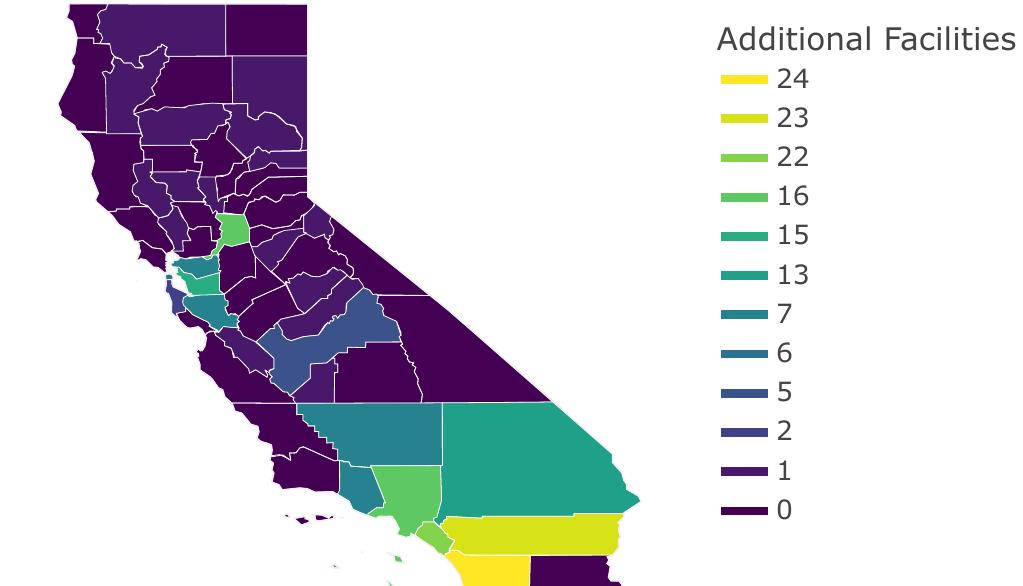}
         \caption{CA Additional Facilities Allocation}
         \label{fig:ca_x}
     \end{subfigure}
        \hfill
     \begin{subfigure}[b]{0.49\textwidth}
         \centering
         \includegraphics[width=\textwidth, trim={6cm 1cm 0.5cm 1cm},clip]{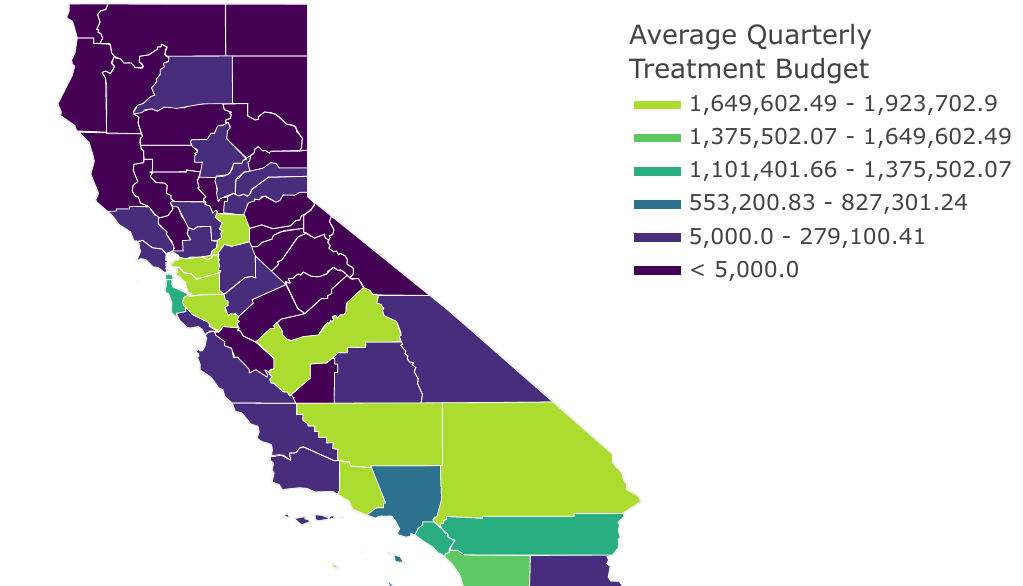}
         \caption{CA Treatment Budget Allocation}
         \label{fig:ca_bud}
     \end{subfigure}


     \bigskip
     \begin{subfigure}[b]{0.49\textwidth}
        \centering
         \includegraphics[width=\textwidth, trim={6cm 0.5 0.5cm 1cm},clip]{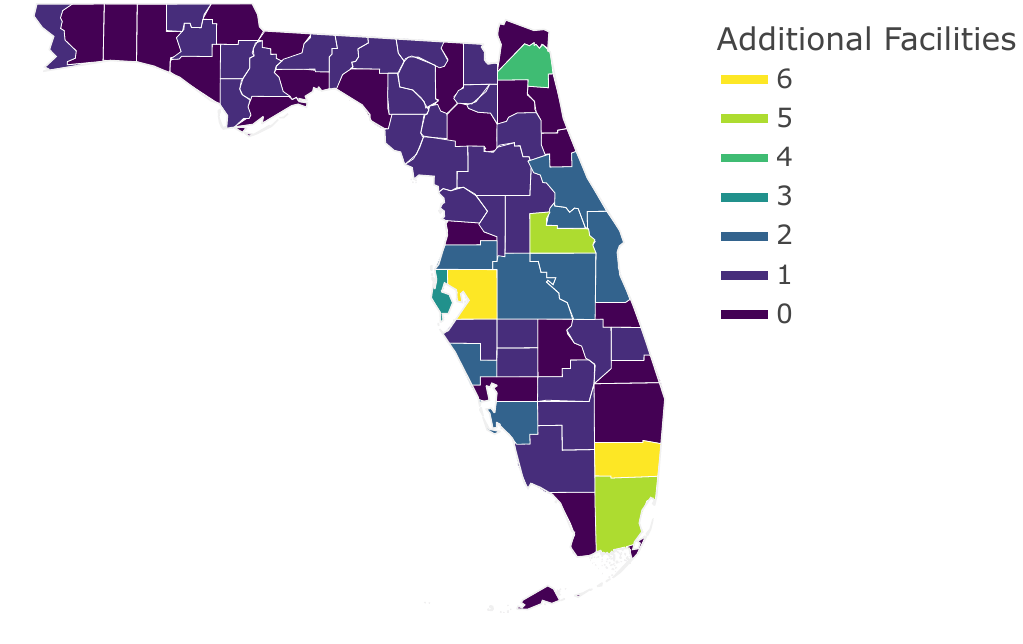}
         \caption{FL Additional Facilities Allocation}
         \label{fig:fl_x}
     \end{subfigure}
        \hfill
     \begin{subfigure}[b]{0.49\textwidth}
         \centering
         \includegraphics[width=\textwidth, trim={6cm 0.5 0.5cm 1cm},clip]{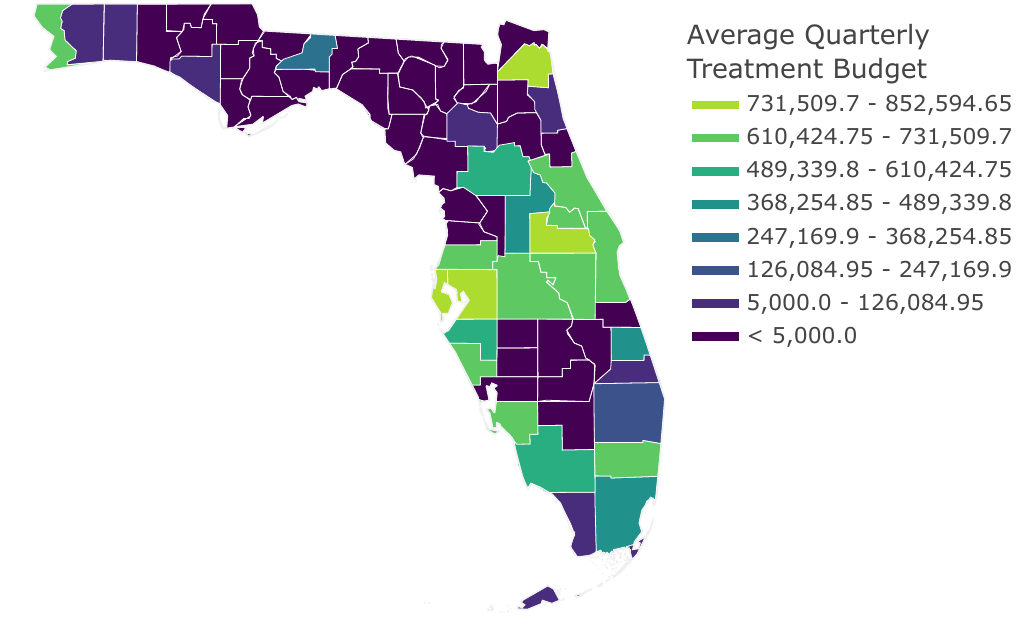}
         \caption{FL Treatment Budget Allocation}
         \label{fig:fl_bud}
     \end{subfigure}

        \caption{State MIP solutions consisting of additional facilities and treatment budget allocations.}
        \label{fig:sols}
\end{figure}
\else
\begin{figure}[htbp]
     \centering

     \begin{subfigure}[b]{0.44\textwidth}
        \centering
         \includegraphics[width=\textwidth, trim={6cm 1cm 0.5cm 1cm},clip]{./Figures/map_x_CA.pdf}
         \caption{CA Additional Facilities Allocation}
         \label{fig:ca_x}
     \end{subfigure}
        \hfill
     \begin{subfigure}[b]{0.44\textwidth}
         \centering
         \includegraphics[width=\textwidth, trim={6cm 1cm 0.5cm 1cm},clip]{./Figures/map_bud_CA.pdf}
         \caption{CA Average Treatment Budget Allocation}
         \label{fig:ca_bud}
     \end{subfigure}


     \bigskip
     \begin{subfigure}[b]{0.44\textwidth}
        \centering
         \includegraphics[width=\textwidth, trim={6cm 0.5 0.5cm 1cm},clip]{./Figures/map_x_FL.pdf}
         \caption{FL Additional Facilities Allocation}
         \label{fig:fl_x}
     \end{subfigure}
        \hfill
     \begin{subfigure}[b]{0.44\textwidth}
         \centering
         \includegraphics[width=\textwidth, trim={6cm 0.5 0.5cm 1cm},clip]{./Figures/map_bud_FL.pdf}
         \caption{FL Average Treatment Budget Allocation}
         \label{fig:fl_bud}
     \end{subfigure}

        \caption{State MIP solutions consisting of additional facilities and average treatment budget allocations.}
        \label{fig:sols}
\end{figure}
\fi

\secondReviewChanges{Figures~\ref{fig:ca_bud} and~\ref{fig:fl_bud} show the average quarterly treatment budget allocations yielded by the MIP solutions for California and Florida. We average over time since certain counties have changing quarterly budget recommendations. We further analyze these trends in \ifpreprint Appendix~\ref{trends}. \else Section EC.5. \fi On average, more of the treatment budget limit per quarter is distributed to the counties in each state which have more recommended additional facilities. This makes sense because we need to ensure that each individual facility has enough treatment capacity to see a certain number of patients per quarter. However, it is noteworthy that budget allocation is not exactly proportional to the facilities allocation. This could be a result of certain counties being over-saturated with treatment facilities, so our MIPs are compensating by allocating the more of the budget to counties that have a moderate amount of facilities but could benefit more from the additional treatment capacity. In counties with numerous facilities, patients are spread across facilities, which reduces the need for larger treatment budgets per facility to achieve similar benefits.}

\paragraph{Solution Impact.}
We quantify the effect of optimizing the locations of additional treatment facilities and the treatment budget on the compartments $A$, $R$, and $D$ after 2 years for almost all US states (full table in \ifpreprint Appendix~\ref{tab_fig}). \else Section EC.11). \fi In comparison to our baseline compartmental model predictions, the proposed solutions to the respective state MIPs on average decrease the number of people with OUD by 
\secondReviewChanges{$9.03\pm1.772\%$}, increase the number of people getting treatment by 
\secondReviewChanges{$88.75\pm26.223\%$}, and decrease the number of opioid-related deaths by 
\secondReviewChanges{$0.58\pm0.111\%$} after 2 years (Figure~\ref{fig:percent_changes}).

\ifpreprint
\begin{figure}[htb]
\centering
\includegraphics[width=.7\textwidth,trim={0.5cm 0.7cm 0.5cm 2cm},clip]{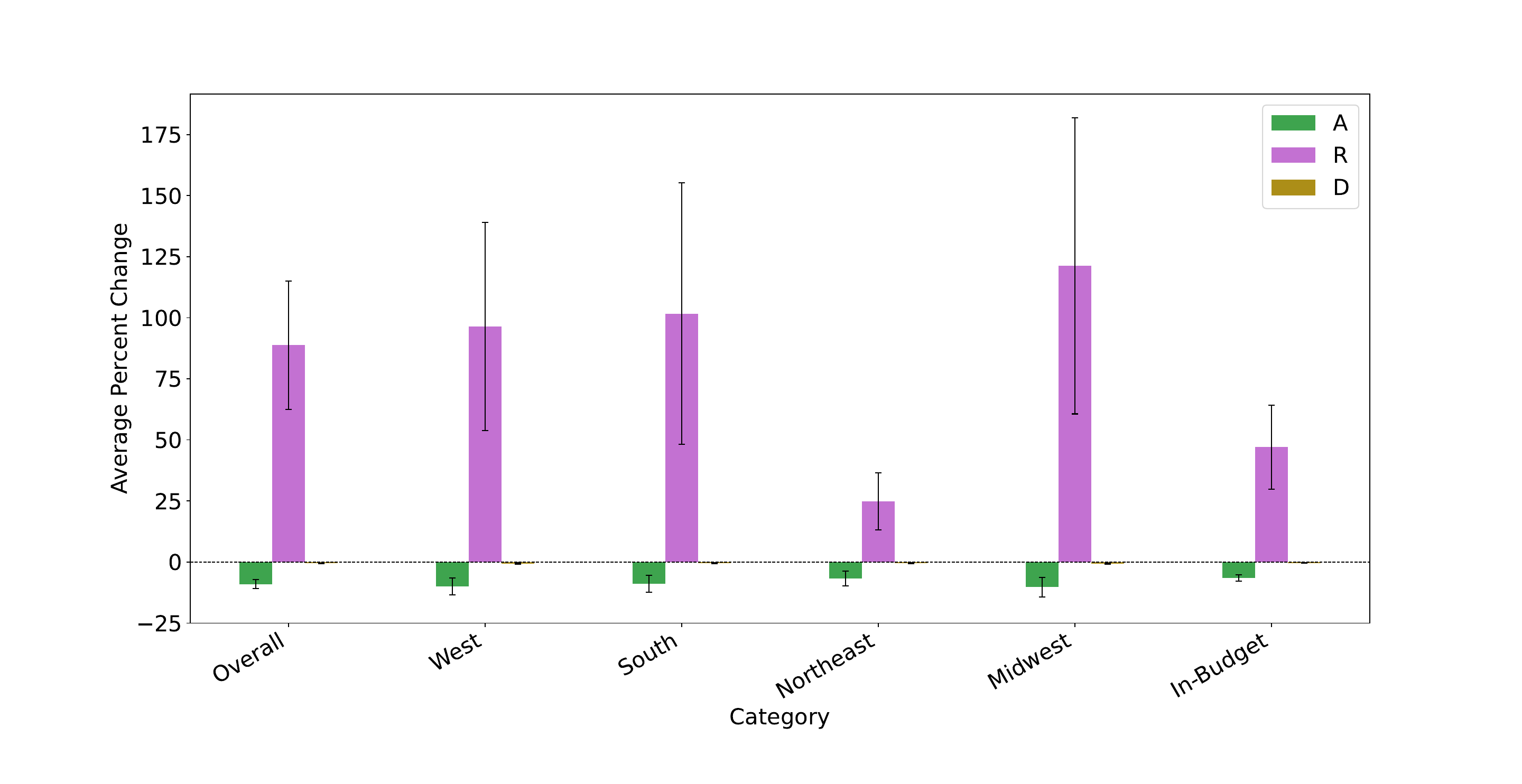}
\caption{Average effect of our MIP solutions on compartments $A$, $R$, and $D$ for various solution groupings. }
\label{fig:percent_changes}
\end{figure}
\else
\fi

\ifpreprint \else
\begin{figure}[htb]
\centering
\includegraphics[width=.7\textwidth,trim={0.5cm 0.7cm 0.5cm 2cm},clip]{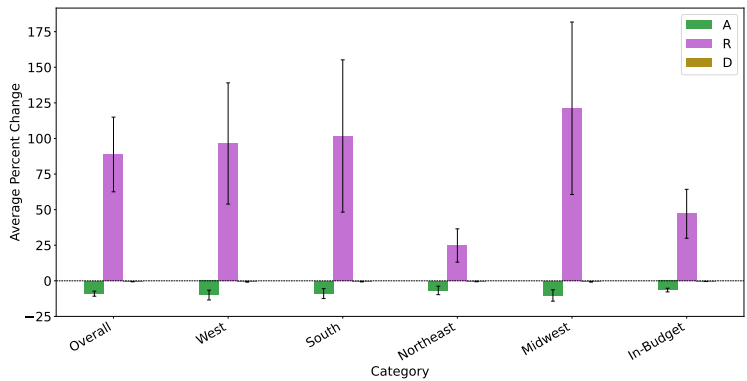}
\caption{Average effect of our MIP solutions on compartments $A$, $R$, and $D$ for various solution groupings. }
\label{fig:percent_changes}
\end{figure}
\fi

Figure~\ref{fig:percent_changes} shows the average effect of the MIP solutions on compartments $A$, $R$, and $D$ for the main US geographic regions: Northeast, West, Midwest, and South. \secondReviewChanges{States in the Midwest have a much larger average effect on the $R$ compartment compared to Northeast states. This is because we allow for overbudget solutions, where $\sum_{i \in \mathcal{C}} x_i > N$, to ensure that each county has at least one treatment facility. Midwest states tend to have overbudget solutions since they have many small counties with no initial treatment facilities. Adding a facility to each of these small counties therefore increases treatment capacity significantly,
leading to states with overbudget solutions having over 100\% increases in the $R$ compartment compared to baseline epidemic dynamics. We show these solutions in \ifpreprint Appendix~\ref{overbud}. \else Section EC.6. \fi If we only consider solutions that are within budget, Figure~\ref{fig:percent_changes} shows that these solutions on average decrease the number of people with OUD by 
\secondReviewChanges{$6.49\pm1.295\%$}, increase the number of people getting treatment by 
\secondReviewChanges{$47.08\pm17.195\%$}, and decrease the number of opioid-related deaths by 
\secondReviewChanges{$0.43\pm0.075\%$} after 2 years. } \thirdReviewChanges{For our formulation with spatial information, there are fewer states with over-budget solutions, as we allow certain counties to have no treatment facilities. However, the spatial formulation solutions still have a similar impact on population health measures compared to the original formulation, as shown in Section~\ref{bench_descrip}. We show a detailed analysis of the spatial formulation solutions in \ifpreprint Appendix~\ref{spatial}. \else Section EC.7. \fi}
Across states, our MIP solutions have the greatest impact on the number of people in rehabilitation ($R$) because our decision variables directly affect the parameter $\zeta$, which dictates how the $R$ compartment evolves over time. \secondReviewChanges{Although our solutions could significantly increase the number of people being treated for OUD, the trickle down effect to decreasing OUD-related deaths is much smaller. 
This makes sense because treatment decreases the likelihood of relapse and overdose, but does not eliminate the possibility.}
\reviewChanges{

\subsection{Sensitivity and Robustness Analysis}
There is significant uncertainty regarding true opioid epidemic dynamics. We therefore assess how sensitive and robust the recommended policies are to uncertainty in the estimated compartmental model parameters for each state.

\paragraph{Sensitivity.}
We vary estimated parameters $\epsilon$, $\mu$, and $\zeta$, and show
the impact on compartments $A, R,$ and $D$ in comparison to baseline 
state dynamics. 
The estimated parameters are perturbed by 20\%, 50\%, and 80\% in the positive and negative directions.
Figure~\ref{fig:param_change} shows percent changes in the $A, R,$ and $D$ values \secondReviewChanges{compared to baseline compartmental model estimates} for different values of $\epsilon$, $\mu$, and $\zeta$ for the California MIP solution. The dashed line in each subplot indicates the percent changes in $A$, $R$, and $D$ with our nominal estimated parameters. As $\epsilon$ increases, there is a greater projected increase in the number of people in rehabilitation and a greater decrease in the number of addicted people, but a slightly smaller decrease in deaths compared to the baseline. As $\mu$ increases, the effect on the values of $A$ and $R$ remains similar, but there is a greater decrease in deaths. 
This makes sense because $\mu$, the death rate of addicts, only affects the value of $D$. 
For $\zeta$, which is the rate of entry into rehabilitation, we see that as $\zeta$ decreases, our solution has a significantly greater impact on the number of people in rehabilitation and the number of deaths. This analysis indicates that as the estimated parameter values get ``worse,'' our solutions yield greater benefits.

\begin{figure}[htb]
\centering
\includegraphics[width=\textwidth,trim={2.5cm 1cm 4cm 2.5cm},clip]{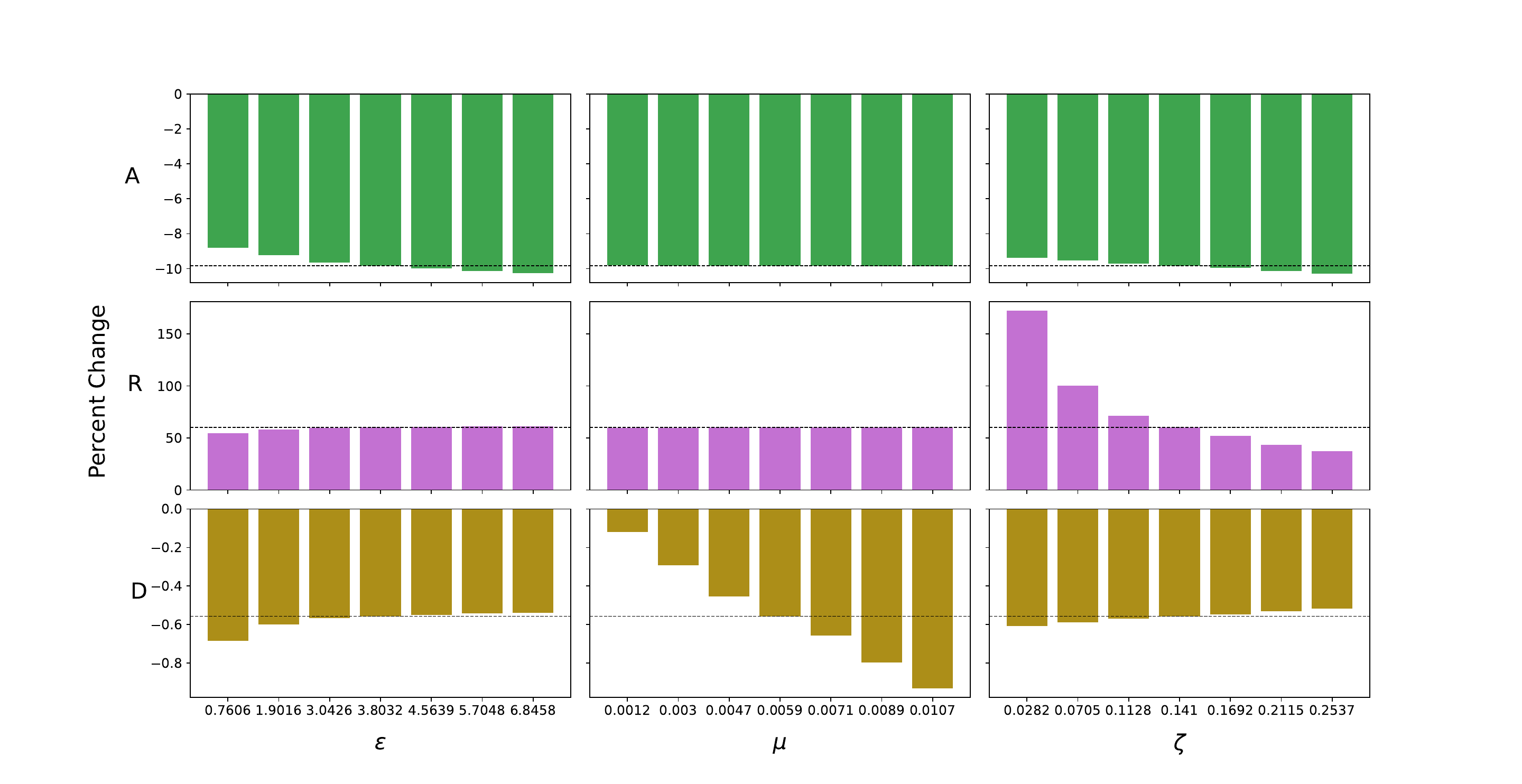}
\caption{Effect of our California MIP solution on the values of compartments $A$, $R$, and $D$ in comparison to baseline opioid epidemic dynamics after 2 years for different compartmental model parameter ranges.}
\label{fig:param_change}
\end{figure}

\secondReviewChanges{
\paragraph{Robustness.}
We assess the robustness of our MIP solutions by re-solving our MIP formulation with the ranges of estimated model parameters used in the sensitivity analysis as inputs.
\thirdReviewChanges{Our results (in \ifpreprint Appendix~\ref{robust1}\else Section EC.8.1\fi) show that the solution differences when varying the individual parameters are minimal for various states, which indicates that our solutions are highly robust to uncertainty in $\epsilon$, $\mu$, and $\zeta$.}
In similar facility location work (\ie, \citet{vax_location}), they also show that their facility location solutions are highly robust to parameter changes. 

\thirdReviewChanges{
We also develop a robust formulation that addresses uncertainty in the population compartment values at each time period. Following~\citet{robust}, we use our original deterministic compartmental model dynamics to determine the nominal compartment values, but then robustify the operational constraints that are dependent on the compartment values. For most states, the robust solution either remains the same as the original solution or leads to additional benefits in terms of the percentage changes in the values of $A$, $R$, and $D$ compared to the baseline compartmental model dynamics.
The formulation and detailed analysis is shown in \ifpreprint Appendix~\ref{robust2}. \else Section EC.8.2. \fi}
}
}
\secondReviewChanges{
\subsection{Comparison with Benchmarks}\label{bench_descrip}
We show 4 different benchmarks where we combine simple methods for deciding on the facility location allocation and the treatment budget allocation. The facility location allocation is either population-based or SVI ranking-based. Then, we either distribute the treatment budget proportionally to the population in each county or uniformly across counties in each time period. We show the baseline optimization problem for the population-based benchmark:
\begin{equation*}\label{eq:baseline_ex}
\begin{array}{ll}
        \text{minimize } & \left\|(\mathbf{Pop}/\sum_{i=1}^C \text{Pop}_i)N - \mathbf{x}\right\|_1 + \lambda_{\mathrm{inf}}\max\left(0, \sum_{i = 1}^C x_i - N\right) \\
	    \text{subject to} &  x_i \geq n_i, \quad  \forall i \in \mathcal{C}\\
        & x_i \in \integers, \quad x_i \geq 1, \quad   \forall i \in \mathcal{C}.
        \end{array}
\end{equation*}
Here we try to make the facilities allocation proportional to the county population. Similarly, for the SVI ranking-based baseline, we simply replace the county population vector $\mathbf{Pop}$ in the optimization problem with the vector of SVI rankings. We show how our method's solutions compare to the simple benchmarks in Table~\ref{tab:baseline_compare}. 

\thirdReviewChanges{\paragraph{Impact.} Our optimized solution has a significantly larger effect on compartments $A$, $R$, and $D$ compared to the benchmarks in which the treatment budget is distributed proportionally to the population (regardless of the facility location method).} For the uniformly-distributed treatment budget benchmarks, the benchmarks still do worse than our optimized model, but the performance is more comparable. This is likely because we are taking into account these exact factors of SVI ranking and population share within our optimization problem. Nevertheless, our MIPs have further flexibility that allows for even better solutions, and in turn, a greater improvement in these population health measures. 
\thirdReviewChanges{If we considered an optimized solution with no socioeconomic considerations,}
this would likely perform even better in terms of the population health measures. However, our optimized solutions that take into account socioeconomic equitability still lead to benefits in comparison to benchmarks that solely consider population or social vulnerability. 
\thirdReviewChanges{From Table~\ref{tab:baseline_compare}, we see 0.76, 3.98, and 0.05 percentage point improvements on average in the effects on compartments $A$, $R$, and $D$ compared to the best demographic-based benchmark (SVI-based facilities, uniformly distributed budget). Depending on the population size of the state, small percentage improvements can correspond to helping many additional people.} \thirdReviewChanges{We also show that our optimized solution with spatial information performs significantly better than the SVI- and population-based benchmarks with spatial information.} We compare map plots of our solutions and benchmarks in \ifpreprint Appendix~\ref{compbench}. \else Section EC.9. \fi

\begin{table*}[h]
\ifpreprint 
\small
\setlength\tabcolsep{0.2em}
\else
  \TABLEsmallX
  \fi
  \centering
\caption{\thirdReviewChanges{Comparison between our optimized solutions and benchmark solutions in terms of facility allocation equitability and average effect on compartments $A$, $R$, and $D$.}}
  \label{tab:baseline_compare} 
  \ifpreprint
  \adjustbox{max width = 1.2\textwidth}{%
  \fi
  \adjustbox{max width=\textwidth}{%
  \begin{tabular*}{0.85\textwidth}{@{\extracolsep{\fill}}llcccc@{\extracolsep{\fill}}}
    \toprule
    Facility Location & Treatment Budget & $A$ & $R$ & $D$ & Equitability Loss \\
    \midrule
    \begin{filecontents*}{./Data/baseline_compare_table_new_display.csv}
Facility Location,Treatment Budget ,A,R,D,Equitability
Population-based,Population-based,-4.41,40.42,-0.31,156.75
Population-based,Uniformly Distributed,-8.25,84.58,-0.52,156.75
Population-based Spatial,Uniformly Distributed,-6.80,70.78,-0.44,161.83
SVI-based ,Population-based,-2.88,29.59,-0.22,141.23
SVI-based ,Uniformly Distributed,-8.27,84.77,-0.53,141.23
SVI-based Spatial,Uniformly Distributed,-7.40,75.94,-0.48,114.36
Optimized,Optimized,-9.03,88.75,-0.58,147.48
Optimized Spatial,Optimized Spatial,-8.40,84.40,-0.54,159.11
\end{filecontents*}
\csvreader[head to column names, late after line=\\]{./Data/baseline_compare_table_new_display.csv}{
    Facility Location = \fl, Treatment Budget = \tb, A = \A, R = \R, D = \D, Equitability = \eq}{\fl & \tb & \A & \R & \D & \eq}
    \bottomrule
  \end{tabular*}
  }
  \ifpreprint
  }
  \fi
    
\end{table*}
}

\thirdReviewChanges{
\paragraph{Equitability.} 
We develop a quantitative measure of how equitable different proposed facility allocations are based on their 1-norm distance from the normalized SVI ranking: $\left\|(\mathbf{SVI}/\sum_{i=1}^C \text{SVI}_i)N - \mathbf{x}\right\|_1.$ Proportionality to SVI rankings gives greater allocations to counties with higher social vulnerability, which we consider to be the most equitable allocation of facilities. Therefore, a smaller value of our measure (\ie,``equitability loss'') represents greater equitability. 
Table~\ref{tab:baseline_compare} shows that on average across states, 
the SVI-based and SVI-based spatial benchmarks are the most equitable, which is expected, as these benchmarks only aim to minimize equitability loss.
Our optimized solution is the next most equitable solution overall.
In some states, our optimized or optimized spatial solutions are the most equitable. For a detailed comparison at the state level, see \ifpreprint Appendix~\ref{tab_fig}. \else Section EC.11. \fi

}

\section{Conclusions}\label{conclu}
In this work, we develop a novel optimization approach that considers complex opioid epidemic dynamics to compute high-quality opioid treatment facility and treatment budget allocations. The integration of a prescriptive MIP with a dynamical model gives us the ability to show the direct impact of the MIP solutions on epidemic dynamics, and helps the MIP yield solutions that maximize positive impact on population health measures described by the epidemic model.

Our compartmental ODE model formulation expands on previous models of the opioid epidemic by additionally including an illicit opioid use compartment and a deceased compartment. This helps us capture illicit opioid use dynamics and incorporate cumulative overdose death data more concretely. Although there have been previous state- and national-level compartmental ODE models of the opioid epidemic defined in the literature, no past work estimates unique parameters for almost every US state. We are able to capture the differences in the dynamics of the epidemic between states and interpret these differences through the model parameters. 

\reviewChanges{We then obtain high-quality resource allocation solutions to our MIPs using our efficient McCormick envelope-based method, which takes around \secondReviewChanges{2} seconds on average \secondReviewChanges{with a mean optimality gap of 3.99\% (10.39\% maximum).}
\thirdReviewChanges{Using a quantitative measure of equitability loss, we show that our optimized solutions are comparably equitable in relation to the social vulnerability index.}
Although the impact differs for each state \secondReviewChanges{and we allow for overbudget solutions}, our proposed solutions on average decrease the number of people with OUD by 
\secondReviewChanges{$9.03\pm1.772\%$}, increase the number of people getting treatment by 
\secondReviewChanges{$88.75\pm26.223\%$}, and decrease the number of opioid-related deaths by 
\secondReviewChanges{$0.58\pm0.111\%$} in comparison to our baseline compartmental model predictions after 2 years.
\secondReviewChanges{
\thirdReviewChanges{Compared to several demographic-based benchmarks, our solutions show additional improvements in the aforementioned population health metrics.}
}
}

\paragraph{Limitations.}
Our approach has limitations. Even though our \secondReviewChanges{neural ODE-inspired model fitting process} can deal with irregularly-sampled time series data, data quality and availability was a challenge when estimating parameters for our compartmental models. Having improved and more refined data related to illicit opioids, particularly fentanyl use, could benefit the quality of the parameter estimation, but we still have established that a neural ODE framework can be used for this application. \secondReviewChanges{As a result of limited data, we also only consider time invariant parameters rather than dynamic parameters that change over time. We hope to address this in future work.} 
\secondReviewChanges{Additionally, our epidemiological model and intervention representations are simplifications of real-world dynamics. Due to aggregate data limitations, subtler social and shorter-term dynamics cannot be captured. For instance, contact with illicit opioid providers affecting illicit use is not addressed. We also assume treatment limitations are solely due to capacity, not patient willingness, which cannot be accounted for with aggregate data. Nevertheless, for broader public health considerations, aggregate data remains useful. In terms of intervention modeling, we currently do not allow the shifting of facility resources between counties, but we aim to address this in future work.}

\reviewChanges{
\paragraph{Generalizability of the Modeling Framework.}

Our integrated approach offers a versatile solution, capable of accommodating various epidemiological model variations, MIP formulations, and interventions. As highlighted in \citet{haffajee}, the HHS has identified four key areas of focus for the coming years: primary prevention, harm reduction, evidence-based treatment, and recovery support. Our current work focuses on evidence-based treatment, but it can be expanded to address other priority areas as well. For example, it can be adapted to tackle harm reduction, involving the distribution of lifesaving naloxone and the prevention of infectious diseases like HIV \citep{haffajee}. The same compartmental model can be used to optimize naloxone distribution; we would just have the intervention affect $\mu$, the death rate of addicts. To combat HIV transmission related to injection drug use, we can use compartmental models for HIV disease spread \citep{hiv_sir} and then optimize safe syringe service program locations within states. Furthermore, our modeling framework can apply to primary prevention by optimizing education efforts surrounding opioid use and assist in optimizing recovery housing or recovery community center locations for recovery support. It can also facilitate combined interventions that affect multiple model parameters.

Moreover, our adaptable modeling framework can be extended to address other drug-related epidemics, such as methamphetamines. Utilizing existing compartmental models \citep{Mushanyuzy2016TheRO}, we can optimize interventions to combat the methamphetamine crisis. Given the lack of established drug-based treatments \citep{c37e243cc1e04213b39b1392330c89f7}, our focus can shift to optimizing educational interventions. In summary, our modeling framework's flexibility enables it to address various epidemics and interventions as long as they can be represented by compartmental models.}

\paragraph{Implications.}
Our contributions are two-fold: (1) we provide interpretable parameters which quantify the differences between the opioid epidemic dynamics of different states through parameter estimation with our \secondReviewChanges{neural ODE-inspired model fitting process}, and (2) we formulate and solve a novel MIP approach that proposes more equitable, high-quality solutions for opioid treatment facility location and treatment budget allocation. We show that the proposed solutions could have a positive impact, even in the short term, on population health measures, \secondReviewChanges{and have a greater impact compared to benchmarks.} \reviewChanges{We also find that policy-makers should target moving or adding treatment facilities to counties that have significantly \fourthReviewChanges{fewer} facilities than their population share and counties that are more socially vulnerable.}
In contrast from previous work, our approach directly provides actionable policies based on real-world data regarding opioid treatment allocation. Combined with easy-to-use graphical visualization tools, this approach could be used by policy-makers to inform decision-making regarding the opioid epidemic in the future.

\ifpreprint
\bibliographystyle{plainnat}
\else
\bibliographystyle{informs2014} 
\fi
\bibliography{bib} 

\ifpreprint \else
\ACKNOWLEDGMENT{%
}
\fi

%
%
%


\ifpreprint
\begin{appendices}
\section{Compartmental Model Initial Condition}\label{ic}
Since data was not available for every compartment in 1999, we calculated most of the initial compartment values for each state based on previous literature. For the $P$ compartment, \citet{initial_P} indicates that the sales of prescription opioids in 2010 were 4 times those in 1999, which we used to compare the number of people using prescription opioids in 1999 and 2010, respectively. We divided the value for $P$ in 2010 by 4 to get the initial $P$ compartment value for each state. For the $I$ compartment, we performed calculations based on data stating that the annual average rate of past-year heroin use was 1.6 per 1000 persons in 2002--2004 nationally compared to 2.6 per 1000 persons in 2011--2013~\citep{initial_I}. We approximated the initial $I$ compartment value by multiplying the value for $I$ in 2016 (the earliest year of data we had for this compartment) by $1.6/2.6$ for each state. We assumed that the illicit users population did not change much from 1999--2002 and 2013--2016. For the $A$ compartment, we utilized data indicating that the prevalence of prescription opioid use disorder was 0.6\% in 2003 compared to 0.9\% in 2013~\citep{10.1001/jama.2015.11859}. We multiplied the value for $A$ in 2016 (the earliest year of data we had for this compartment) by $6/9$ for each state. This resulting value was used as the initial $A$ compartment value, as we assumed the addicted population did not change much from 1999--2003 and 2013--2016. For the $R$ compartment, we multiplied the data from 2000 by 0.75 to get the initial value for each state. The multiplier was determined based on the data trend. For the $D$ compartment, we had available data for each state from 1999. The initial $S$ compartment value was calculated based on data from the other 5 compartments and the populations of each US state in 1999.

\section{Reformulation of State-Level MIP}\label{state_reform}
We reformulate the original problem to have at most quadratic constraints and a linear objective. We add new decision variables $z_i$ for all $i \in C$. 
The $z_i$'s are continuous variables that are defined to turn the constraint where we divide two decision variables into a quadratic constraint. 
We have the expression $\bar{d}_{ik}/dx_i$, and we define $z_i = 1/x_i$ by adding the constraint $x_i \cdot z_i = 1$ for all $i \in \mathcal{C}$. 
We reformulate the expression as $\bar{d}_{ik}z_i/d$. We bound $x_i$ such that $x_i \in [1, N]$. Each county should have at least one treatment facility offering MAT, and $N$ is the maximum number of facilities that can be opened within the state. We also define
a variable $h$ to linearize the final objective term. 

\reviewChanges{We also scale variables to prevent 
numerical issues. We introduce new variables $\tilde{d}_{ik}$ and let $\bar{d}_{ik} = \tilde{d}_{ik}\cdot 100000$. We also introduce $\tilde{S}_k, \tilde{P}_k, \tilde{I}_k, \tilde{A}_k, \tilde{R}_k, \tilde{D}_k $ and let $S_k = \tilde{S}_k\cdot f, P_k = \tilde{P}_k\cdot f$, etc., where $f = 100$ if $D_{K} < 10000$ and $f = 1000$, otherwise. We can write the resulting problem as follows:
\secondReviewChanges{
\begin{align*}
    \text{minimize}\:\: & \lambda_D D_K + \lambda_A\sum_{k = 0}^K (A_k) + \lambda_{\mathrm{pr}}\mathbf{c}_{\mathrm{pr}}^{\top}\mathbf{x} + \lambda_{\mathrm{SVI}}\mathbf{c}_{\mathrm{SVI}}^{\top}\mathbf{x} + \lambda_{\mathrm{Pop}}\sum_{k = 0}^K \mathbf{c}_{\mathrm{Pop}}^{\top}\mathbf{\bar{d}}_{k}+ \lambda_{\mathrm{inf}}h  \\
		\text{subject to}\:\: & \sum_{i = 1}^C x_i - N \leq h, \quad 0 \le h  \\
		& x_i \geq n_i, \quad  x_i \cdot z_i = 1,\quad \forall i \in \mathcal{C} \\
		& \sum_{i=1}^C \bar{d}_{ik} \leq d_k, \:  \forall k \in \mathcal{K} \\
        & \bar{d}_{ik}z_i \geq d_{\mathrm{min}}, \quad
            \forall i \in \mathcal{C}, k \in \mathcal{K} \\
         & \bar{d}_{i,k-1} - \bar{d}_{ik} \leq \theta , \quad
    \forall i \in \mathcal{C}, k \in 1, \dots, K \\
		& \mathrm{lb}_{ik} \leq \bar{d}_{ik} \leq \mathrm{ub}_{ik}, \quad
        \forall i \in \mathcal{C},k \in \mathcal{K}\\
		&S_{k+1} = S_k + (\epsilon P_k + \delta R_k - \alpha S_k)\Delta, \quad k=0,\dots,K-1\\
        &P_{k+1} = P_k + (\alpha S_k - (\epsilon + \gamma + \beta)P_k)\Delta, \quad k=0,\dots,K-1\\
        &I_{k+1} = I_k + (\beta P_k - \phi I_k)\Delta, \quad k=0,\dots,K-1\\
        &A_{k+1} = A_k + (\gamma P_k + \sigma R_k + \phi I_k - \left(\zeta + \ell_k\right) A_k - \mu A_k)\Delta, \quad k=0,\dots,K-1\\
        &R_{k+1} = R_k + \left(\left(\zeta + \ell_k\right) A_k - (\delta + \sigma)R_k\right)\Delta, \quad k=0,\dots,K-1\\
        &D_{k+1} = D_k + (\mu A_k)\Delta, \quad   k=0,\dots,K-1\\
        &x_i \in \integers, \quad x_i \in [1, N], \quad \forall i \in \mathcal{C}\\
        &\bar{d}_{ik}, u_i, v_i, z_i, w_k, S_k, P_k, I_k, A_k, R_k, D_k  \geq 0, \quad \forall i \in \mathcal{C},k \in \mathcal{K},
\end{align*}
where we define
$$\ell_k = \frac{\sum_{i = 1}^C d^{-1}(x_i - n_i)\bar{d}_{ik}z_i}{A_k},$$
and 
$$\mathrm{lb}_{ik} = d_{\mathrm{min}} n_i \text{ if } n_i \geq 1, \quad d_{\mathrm{min}} \text{ otherwise}; \quad \mathrm{ub}_{ik} = d_{\mathrm{max}}(n_i + a). $$
We set $d_{\mathrm{max}}$ to a large enough value to ensure that the optimization solution can use the entire treatment budget limit $d_k$ for that period, but we also wanted this bound to be as tight as possible. This involved checking whether the solution for each state was allocating the entire treatment budget limit $d_k$ in the first period, and if not, we then set $d_{\mathrm{max}}$ to be a larger value. We set $d_{\mathrm{max}} = 14000$ for Delaware,  $d_{\mathrm{max}} = 32000$ for Hawaii, $d_{\mathrm{max}} = 12000$ for Nevada, $d_{\mathrm{max}} = 5000$ for Rhode Island, and $d_{\mathrm{max}} = 10000$ for all other states.}}
\reviewChanges{This mixed-integer non-convex bilinear optimization problem can be solved by Gurobi only for small instances. We show our strong relaxation that provides high-quality solutions for all problem instances in Appendix~\ref{mcc}.}

\reviewChanges{\section{McCormick Envelope-Based Relaxation}\label{mcc}
We use McCormick envelopes to replace each bilinear term in our original formulation with its concave and convex envelopes~\citep{mccormick}. 
We have two sets of bilinear terms that appear in the following constraints:
\begin{align*}
& x_i \cdot z_i = 1,\quad \forall i \in \mathcal{C} \\
& A_{k+1} = A_k + (\gamma P_k + \sigma R_k + \phi I_k - \left(\zeta + \ell_k\right) A_k - \mu A_k)\Delta, \quad k=0,\dots,K-1\\
& R_{k+1} = R_k + \left(\left(\zeta + \ell_k\right) A_k - (\delta + \sigma)R_k\right)\Delta, \quad k=0,\dots,K-1,
\end{align*}
where
$$\ell_k = \frac{\sum_{i = 1}^C d^{-1}(x_i - n_i)\bar{d}_{ik}z_i}{A_k}.$$
We have a set of integer $\times$ continuous terms $x_i\cdot z_i \:\:\forall i \in \mathcal{C}$ as well as a set of continuous $\times$ continuous terms $\bar{d}_{ik}\cdot z_i \:\:\forall i \in \mathcal{C}, k \in \mathcal{K}$. 
\paragraph{Nonnegative integer times continuous (exact reformulation).}
For each non-negative integer and continuous product, we can first replace the integer variable with its unary expansion and then use McCormick envelopes to linearize the resulting product of continuous and binary variables~\citep{bincont}. 
This is an exact reformulation, not a relaxation. 
We have integer variables $x_i \in [1, N]$ and continuous variables $z_i \in [1/N, 1]$ in the first set of bilinear terms. 
Let $p_{ij}$ be a new set of binary variables with $j = 1, \dots, N$, and define variables $o_{ij}$ to model the products $o_{ij} = z_i\cdot p_{ij}$. 

We obtain:
\begin{align*}
& x_i = \sum_{j=1}^N j\cdot p_{ij}, \quad \sum_{j=1}^N p_{ij} \leq 1, \quad \sum_{j=1}^N j\cdot o_{ij} = 1 \quad \forall i \in \mathcal{C} \\
& o_{ij} \geq \frac{1}{N} p_{ij} \quad \forall i\in \mathcal{C}, j = 1, \dots, N \\
& o_{ij} \geq p_{ij} + z_i - 1 \quad \forall i\in \mathcal{C}, j = 1, \dots, N\\
& o_{ij} \leq  p_{ij} \quad \forall i\in \mathcal{C}, j = 1, \dots, N\\
& o_{ij} \leq  z_i + \frac{1}{N}p_{ij} - \frac{1}{N} \quad \forall i\in \mathcal{C}, j = 1, \dots, N.
\end{align*}

\paragraph{Continuous times continuous (relaxation).}
For the product of continuous variables, we can also use McCormick envelopes to relax these terms. 
We have variables $\bar{d}_{ik} \in [\mathrm{lb}_{ik}, \mathrm{ub}_{ik}]$ and $z_i \in [1/N, 1]$. Let $r_{ik}$ be a new set of variables to model the products $r_{ik} = \bar{d}_{ik}\cdot z_i$. 
We use McCormick envelopes to obtain the following relaxation:
\begin{align*}
& r_{ik} \geq \mathrm{lb}_{ik}\cdot z_i + \frac{1}{N}\bar{d}_{ik} - \frac{1}{N}\mathrm{lb}_{ik} \quad \forall i\in \mathcal{C}, k \in \mathcal{K}\\
& r_{ik} \geq \mathrm{ub}_{ik}\cdot z_i + \bar{d}_{ik} -\mathrm{ub}_{ik} \quad \forall i\in \mathcal{C}, k \in \mathcal{K}\\
& r_{ik} \leq \mathrm{ub}_{ik}\cdot z_i + \frac{1}{N}\bar{d}_{ik} - \frac{1}{N}\mathrm{ub}_{ik} \quad \forall i\in \mathcal{C}, k \in \mathcal{K}\\
& r_{ik} \leq  \bar{d}_{ik} + \mathrm{lb}_{ik}\cdot z_i - \mathrm{lb}_{ik} \quad \forall i\in \mathcal{C}, k \in \mathcal{K}.
\end{align*}
Solving the full relaxed problem provides a lower bound on the optimal objective value and a feasible solution to the original problem, which we use to obtain a corresponding upper bound on the optimal objective value (as described in Section 3.3).
}


\reviewChanges{
\section{Epidemiological Model Validation}\label{epi_val}
\secondReviewChanges{We show the performance of our epidemiological model using our estimated parameters (from the neural ODE-inspired model fitting process) by calculating the mean absolute percentage error (MAPE) and the absolute percentage error in comparison to the entire population (pMAPE) for each epidemiological model compartment. Table~\ref{tab:avg_errors} shows these errors for every state and compartment value. We also calculate pMAPE because certain categories like the Deceased category are very small in magnitude compared to the entire population. The population for each state is at least on the order of millions, which means that prediction errors between 100 and 200 individuals (leading to a percent error of 100\%) is not as significant as errors between 100,000 and 200,000. We note that the states Vermont and New Hampshire have very large MAPEs ($\sim 2000\%$) for the $R$ compartment, but in comparison to the population, the pMAPE is only $\sim 0.1\%$. However, overall, our model performs well across states and compartment values in terms of the mean and median MAPE across states (Table~\ref{tab:sum_compare}). Removing Vermont and New Hampshire improves the average MAPE across states, in particular for the $R$ compartment. Table~\ref{tab:sum_compare} shows this improvement is from an average of 133.19\% to 31.65\%, which means that all other states have good performance. 
}

\secondReviewChanges{We also test our model performance using parameters from the literature, specifically from \citet{battista} for $(\phi, \epsilon, \beta, \zeta, \mu)$. \citet{battista}'s model parameters are on the national level, but to the authors' knowledge, there are no other papers that estimate model parameters on the state-level, so we use \citet{battista}'s parameters as the closest comparison. For the parameter ranges, we choose the midpoint of the range. They did not have the exact parameters for $\phi$ and $\beta$ in their models. For $\beta$, we summed their parameters for illicit addiction rate based on P-class and A-class as an approximation. For $\phi$, we took the average of our estimated parameters, so it would at least be comparable to our model. We use the following parameters for the comparison: (0.02, 4.4, 0.0036, 1.1, 0.01159). Table~\ref{tab:lit_errors} shows the MAPE for each compartment and state when using parameters from the literature. In Table~\ref{tab:sum_compare}, we compare the median MAPEs (since we have 2 outliers) across states for each compartments. We see that our estimated parameters perform considerably better compared to the parameters from the literature in terms of the median MAPE (as well as the other summary statistics) for all of the compartments. In particular, even when including Vermont and New Hampshire, the average MAPEs are still better when using the estimated parameters versus the literature parameters.}

\begin{table*}[h]
\ifpreprint \small
\else
  \TABLEsmallX
  \fi
  \centering
  \caption{Mean absolute percentage error (MAPE) and mean absolute percentage error compared to the population (pMAPE) for each compartment using neural ODE estimated parameters.}
  \label{tab:avg_errors} 
  \ifpreprint
  \adjustbox{max totalheight=0.9\textheight}{%
  \fi
  \adjustbox{max width=\textwidth}{%
  \begin{tabular*}{1.2\textwidth}{@{\extracolsep{\fill}}lllllllllllll@{\extracolsep{\fill}}}
    \toprule
    & \multicolumn{2}{c}{$S$} & \multicolumn{2}{c}{$P$}& \multicolumn{2}{c}{$I$}& \multicolumn{2}{c}{$A$}& \multicolumn{2}{c}{$R$} & \multicolumn{2}{c}{$D$} \\
    \midrule
    State &   \tiny{MAPE}   &  \tiny{pMAPE} & \tiny{MAPE}  & \tiny{pMAPE}  & \tiny{MAPE} & \tiny{pMAPE} & \tiny{MAPE} & \tiny{pMAPE} & \tiny{MAPE} & \tiny{pMAPE} & \tiny{MAPE} & \tiny{pMAPE} \\
    \midrule
    \begin{filecontents*}{./Data/average_weighting_errors_display.csv}
State,S_MAPE,S_pMAPE,P_MAPE,P_pMAPE,I_MAPE,I_pMAPE,A_MAPE,A_pMAPE,R_MAPE,R_pMAPE,D_MAPE,D_pMAPE
AL,0.96,0.84,12.221,1.32,5.708,0.014,40.246,0.42,11.091,0.013,36.866,0.0048
AZ,0.22,0.20,6.890,0.45,17.550,0.045,57.389,0.51,28.410,0.034,23.246,0.0069
AR,2.23,1.98,23.860,2.27,7.302,0.013,72.989,0.43,46.904,0.009,25.834,0.0021
CA,0.07,0.07,6.790,0.29,8.623,0.015,26.034,0.14,8.028,0.007,7.871,0.0016
CO,0.14,0.13,8.702,0.48,6.438,0.017,52.599,0.42,31.182,0.012,18.933,0.0051
CT,0.15,0.14,7.186,0.36,30.884,0.124,46.305,0.29,17.692,0.054,36.699,0.0196
DE,0.41,0.37,5.652,0.40,10.632,0.062,30.879,0.43,41.729,0.100,38.121,0.0162
FL,0.16,0.15,6.057,0.37,7.169,0.016,21.828,0.17,12.712,0.009,7.604,0.0031
GA,0.97,0.90,20.561,1.38,6.155,0.009,70.601,0.37,36.230,0.018,36.156,0.0057
HI,0.08,0.07,6.891,0.27,0.917,0.002,21.106,0.10,9.918,0.005,4.188,0.0013
IL,0.28,0.26,7.629,0.38,2.666,0.006,34.137,0.17,10.502,0.011,63.186,0.0130
IN,0.22,0.20,5.253,0.43,16.568,0.060,35.015,0.32,49.555,0.021,79.269,0.0101
IA,0.12,0.11,12.502,0.63,6.732,0.013,30.228,0.17,44.403,0.008,41.059,0.0031
KS,0.89,0.82,19.175,1.30,20.126,0.055,46.704,0.23,36.148,0.012,14.112,0.0010
KY,1.06,0.95,6.562,0.60,5.759,0.020,43.381,0.64,77.216,0.041,49.084,0.0166
LA,1.23,1.11,21.900,1.90,8.869,0.017,62.433,0.45,14.384,0.009,29.497,0.0042
ME,0.82,0.76,8.258,0.52,12.249,0.056,20.538,0.20,84.677,0.050,19.099,0.0094
MD,0.12,0.11,5.520,0.29,29.298,0.084,43.593,0.23,23.616,0.070,142.03,0.0263
MA,0.62,0.58,10.416,0.49,11.091,0.040,27.947,0.25,20.783,0.047,81.398,0.0212
MI,0.76,0.69,11.600,0.88,11.832,0.027,61.777,0.32,13.501,0.013,55.160,0.0136
MN,0.30,0.28,17.327,0.75,7.250,0.017,102.32,0.25,44.827,0.023,47.432,0.0054
MO,0.66,0.60,13.958,1.00,7.778,0.019,73.631,0.43,29.667,0.013,35.886,0.0105
NE,0.86,0.81,21.362,1.16,11.664,0.021,62.172,0.25,48.698,0.008,6.695,0.0006
NV,0.48,0.44,11.295,0.83,20.879,0.050,33.775,0.32,27.912,0.015,6.716,0.0045
NH,0.45,0.41,8.761,0.52,31.534,0.135,55.438,0.53,2122.9,0.076,65.507,0.0255
NJ,0.08,0.08,5.706,0.25,10.782,0.045,8.287,0.08,9.258,0.013,23.325,0.0126
NM,0.06,0.06,4.821,0.27,8.128,0.021,20.818,0.12,35.046,0.051,4.174,0.0046
NY,0.04,0.03,4.304,0.15,11.652,0.034,10.436,0.05,10.199,0.022,80.145,0.0135
NC,0.26,0.24,8.099,0.59,14.686,0.041,43.680,0.32,44.886,0.034,16.671,0.0067
OH,0.82,0.75,6.741,0.47,9.482,0.031,36.816,0.39,59.882,0.028,53.236,0.0209
OK,0.68,0.61,14.559,1.33,10.949,0.031,45.241,0.25,58.194,0.016,10.892,0.0053
OR,0.27,0.25,6.768,0.52,8.207,0.029,26.642,0.27,15.414,0.016,12.251,0.0033
PA,0.10,0.09,6.739,0.41,13.723,0.056,23.961,0.18,18.957,0.021,37.166,0.0125
RI,0.46,0.43,9.284,0.53,9.025,0.029,35.504,0.24,17.797,0.064,76.316,0.0235
SC,1.08,0.98,18.350,1.43,7.408,0.013,116.61,0.51,34.850,0.016,40.326,0.0097
TN,0.49,0.44,16.253,1.61,7.301,0.018,58.487,0.53,32.244,0.018,45.601,0.0117
TX,1.05,0.99,33.794,1.79,16.336,0.027,112.29,0.35,8.532,0.003,6.127,0.0008
UT,0.13,0.12,7.462,0.50,7.958,0.020,58.114,0.36,44.348,0.039,6.465,0.0030
VT,0.15,0.14,10.554,0.44,13.190,0.067,39.620,0.47,2204.8,0.145,13.335,0.0064
VA,0.19,0.18,6.121,0.35,11.009,0.028,23.953,0.18,28.425,0.015,12.533,0.0057
WA,0.33,0.30,7.550,0.46,11.851,0.037,32.421,0.32,33.530,0.031,9.290,0.0024
WI,0.18,0.17,5.713,0.32,6.900,0.021,37.274,0.23,44.765,0.020,34.803,0.0098
\end{filecontents*}
\csvreader[head to column names, late after line=\\]{./Data/average_weighting_errors_display.csv}{
    State=\state,
    S_MAPE=\smape, S_pMAPE=\spmape, P_MAPE = \pmape, P_pMAPE = \ppmape, I_MAPE = \imape, I_pMAPE = \ipmape, A_MAPE = \amape,	A_pMAPE =\apmape, R_MAPE = \rmape,	R_pMAPE = \rpmape, D_MAPE = \dmape,	D_pMAPE = \dpmape }{\state & \smape & \spmape & \pmape & \ppmape & \imape & \ipmape & \amape & \apmape & \rmape & \rpmape & \dmape & \dpmape}
    \bottomrule
  \end{tabular*}
  }
  \ifpreprint
  }
  \fi
    
\end{table*}
\clearpage

\begin{table*}[h]
\ifpreprint \small
\else
  \TABLEsmallX
  \fi
  \centering
  \caption{Mean absolute percentage error (MAPE) and mean absolute percentage error compared to the population (pMAPE) for each compartment using literature parameters.}
  \label{tab:lit_errors} 
  \ifpreprint
  \adjustbox{max totalheight=0.9\textheight}{%
  \fi
  {%
  \begin{tabular*}{1.3\textwidth}{@{\extracolsep{\fill}}lllllllllllll@{\extracolsep{\fill}}}
    \toprule
    & \multicolumn{2}{c}{$S$} & \multicolumn{2}{c}{$P$}& \multicolumn{2}{c}{$I$}& \multicolumn{2}{c}{$A$}& \multicolumn{2}{c}{$R$} & \multicolumn{2}{c}{$D$} \\
    \midrule
    State &   \tiny{MAPE}   &  \tiny{pMAPE} & \tiny{MAPE}  & \tiny{pMAPE}  & \tiny{MAPE} & \tiny{pMAPE} & \tiny{MAPE} & \tiny{pMAPE} & \tiny{MAPE} & \tiny{pMAPE} & \tiny{MAPE} & \tiny{pMAPE} \\
    \midrule
    \begin{filecontents*}{./Data/lit_mdpt_errors_display.csv}
State,S_MAPE,S_pMAPE,P_MAPE,P_pMAPE,I_MAPE,I_pMAPE,A_MAPE,A_pMAPE,R_MAPE,R_pMAPE,D_MAPE,D_pMAPE
AL,5.8062,5.1105,62.5497,6.3400,5.9639,0.0145,53.2493,0.5784,176.7438,0.2450,18.8200,0.0058
AZ,2.0960,1.9369,43.1641,2.6416,27.7165,0.0679,40.6270,0.3922,422.2616,0.3839,47.2996,0.0274
AR,5.3436,4.7482,58.9151,5.3943,8.9849,0.0151,50.4527,0.3856,1123.2001,0.2054,60.8164,0.0279
CA,0.5237,0.4986,17.0884,0.7097,19.1916,0.0312,43.6600,0.2561,184.9654,0.1596,57.8464,0.0239
CO,1.4328,1.3410,34.2734,1.8010,7.6979,0.0190,44.2378,0.3757,940.1606,0.3354,54.0101,0.0280
CT,1.7477,1.6240,32.3661,1.6468,29.2944,0.1244,34.6150,0.3430,40.9021,0.1310,59.0626,0.0436
DE,3.4935,3.1601,49.6106,3.4810,16.6254,0.1166,54.5470,0.8052,126.9100,0.2568,50.2334,0.0373
FL,2.0733,1.9276,41.7004,2.4953,7.5500,0.0159,53.4876,0.4804,295.5151,0.2073,70.5957,0.0476
GA,2.6400,2.4363,46.3456,3.0595,28.5375,0.0389,39.6142,0.2288,312.3227,0.1971,54.7497,0.0202
HI,0.3483,0.3329,8.9201,0.3416,4.4266,0.0094,47.4422,0.2462,220.2942,0.1134,70.5164,0.0286
IL,1.4288,1.3424,28.1708,1.3600,3.2541,0.0068,42.7405,0.2594,143.9994,0.1463,59.9053,0.0253
IN,3.4123,3.1086,52.8305,4.0362,20.0121,0.0852,53.8871,0.5020,295.8147,0.2031,38.5363,0.0167
IA,1.7484,1.6381,31.5726,1.6732,6.3484,0.0126,56.4426,0.3896,877.1049,0.1325,64.8257,0.0200
KS,2.5978,2.4030,45.6462,2.9999,12.9292,0.0467,48.8049,0.2556,266.7812,0.1106,69.5540,0.0256
KY,4.5169,4.0354,60.0150,5.5463,5.3584,0.0197,55.5127,0.8363,860.0487,0.4202,58.1121,0.0530
LA,4.1100,3.7102,56.6789,4.8059,9.2412,0.0178,47.1865,0.3984,367.5538,0.2553,41.4823,0.0127
ME,2.4063,2.2263,44.2935,2.8229,29.0170,0.1477,61.0242,0.7374,53.9187,0.0649,76.0707,0.0607
MD,1.7979,1.6734,32.9293,1.6909,30.8640,0.0861,32.5975,0.2368,39.6169,0.1120,61.4210,0.0500
MA,1.1685,1.0992,27.3329,1.3254,19.8246,0.0763,53.2458,0.5540,68.5589,0.1372,59.9054,0.0408
MI,3.3678,3.0723,49.6646,3.5502,12.1212,0.0284,43.2837,0.3562,235.6695,0.2057,63.6507,0.0379
MN,0.5994,0.5701,20.3395,0.8799,8.2854,0.0200,31.6061,0.1264,247.4331,0.1424,55.8931,0.0172
MO,2.9373,2.6943,47.8427,3.2714,10.2040,0.0238,41.3455,0.3242,560.8497,0.2671,65.1964,0.0445
NE,1.6290,1.5283,34.2422,1.8284,17.6086,0.0284,44.0374,0.2078,689.0307,0.1300,59.1760,0.0112
NV,3.1093,2.8372,49.3011,3.4554,33.0271,0.0719,50.3376,0.5559,662.7668,0.3521,67.0364,0.0758
NH,2.2814,2.0998,41.0947,2.4465,32.3280,0.1356,46.4133,0.5615,2056.5768,0.3143,58.7990,0.0571
NJ,1.2228,1.1495,24.1436,1.0868,14.9626,0.0687,53.6914,0.5070,116.1375,0.1647,61.8260,0.0342
NM,2.0523,1.9087,37.6735,2.0892,9.0319,0.0228,44.0888,0.2751,98.2976,0.1156,74.3169,0.0957
NY,0.5573,0.5299,9.3534,0.3506,15.0322,0.0414,42.9340,0.2481,26.4363,0.0557,60.8596,0.0289
NC,3.1365,2.8678,48.8785,3.4253,16.3325,0.0437,45.8212,0.3614,229.9718,0.2127,69.2535,0.0503
OH,2.7684,2.5480,48.7490,3.3935,16.0848,0.0611,60.7802,0.7360,483.8899,0.1965,71.3622,0.0589
OK,4.2180,3.8089,57.7846,5.0604,11.2022,0.0342,53.2644,0.3736,320.5506,0.1217,79.5671,0.0775
OR,2.7033,2.4835,49.6738,3.5190,7.8512,0.0292,51.4984,0.5252,266.7706,0.2708,62.5942,0.0438
PA,2.3606,2.1847,40.8542,2.4255,22.6890,0.1036,52.7700,0.4538,114.1462,0.1392,60.4378,0.0318
RI,1.8241,1.6970,39.6658,2.3119,14.1298,0.0518,37.9379,0.3108,38.9680,0.1261,64.3094,0.0523
SC,3.5045,3.1870,51.7851,3.8898,30.4537,0.0488,39.7080,0.3274,382.6346,0.2826,48.8609,0.0220
TN,4.7395,4.2361,61.4821,5.9625,8.0480,0.0212,50.1795,0.4734,470.9296,0.2892,59.5632,0.0446
TX,1.3166,1.2384,34.4821,1.8212,30.9816,0.0446,33.4321,0.1747,532.9770,0.2218,56.3395,0.0184
UT,2.4607,2.2697,44.9607,2.8831,13.4127,0.0311,40.3753,0.3630,352.7013,0.2803,70.4342,0.0796
VT,1.9952,1.8511,22.0953,1.0037,24.9589,0.1343,55.5970,0.7016,1335.6157,0.2407,68.8837,0.0553
VA,1.7706,1.6534,37.0935,2.0578,10.9198,0.0256,52.4974,0.4217,313.6237,0.1840,68.3786,0.0401
WA,2.0632,1.9103,41.7470,2.5012,13.3385,0.0386,47.6619,0.5091,367.8624,0.3363,62.2411,0.0478
WI,1.7145,1.6031,35.9746,1.9537,15.3733,0.0482,48.6051,0.3563,363.3709,0.1882,70.0366,0.0434
\end{filecontents*}
\csvreader[head to column names, late after line=\\]{./Data/lit_mdpt_errors_display.csv}{
    State=\state,
    S_MAPE=\smape, S_pMAPE=\spmape, P_MAPE = \pmape, P_pMAPE = \ppmape, I_MAPE = \imape, I_pMAPE = \ipmape, A_MAPE = \amape,	A_pMAPE =\apmape, R_MAPE = \rmape,	R_pMAPE = \rpmape, D_MAPE = \dmape,	D_pMAPE = \dpmape }{\state & \smape & \spmape & \pmape & \ppmape & \imape & \ipmape & \amape & \apmape & \rmape & \rpmape & \dmape & \dpmape}
    \bottomrule
  \end{tabular*}
  }
  \ifpreprint
  }
  \fi
    
\end{table*}
\clearpage

}

\begin{table}[htbp]
\ifpreprint 
\centering
\else 
\TABLEsmallX
\fi
\caption{Summary statistics for the overall mean absolute percentage error (MAPE) across all states for each compartment to compare model performance with estimated parameters versus literature parameters.} \label{tab:sum_compare}
\ifpreprint
\centering
\adjustbox{max width=0.7\textwidth}{%
\fi
\centering 
\begin{tabular}{llcccccc}
\toprule
& Metric & $S$ & $P$ & $I$ & $A$ & $R$ & $D$ \\
\midrule
 \multirow{6}{*}{Estimated} & Mean & 0.491 &	10.93	& 11.53 &	45.31	& 133.19 & 34.63 \\
 & Std	& 0.446	& 6.44	& 6.77 & 24.52	& 460.02	& 28.53 \\
& Mean w/o VT, NH &	0.501 & 10.996 &	10.99 & 45.20 & 31.65 & 34.39 \\
& Std w/o VT, NH	& 0.454	& 6.59	 & 6.13	& 25.07	& 18.72	& 28.63\\
& \textbf{Median} & \textbf{0.312}	& \textbf{8.18}	& \textbf{10.06} &	\textbf{39.93}	& \textbf{31.71} & \textbf{32.15}\\
& IQR	& 0.654	& 6.86	& 5.65&	29.41	& 28.83	& 34.65 \\
\midrule
\multirow{6}{*}{Literature} & Mean	& 2.45	& 40.55	& 16.22	& 47.17	& 406.76	& 60.78 \\
& Std	& 1.27	& 13.37 &	8.86	& 7.37	& 398.92	& 10.87 \\
& Mean w/o VT, NH	& 2.47	& 41.00	& 15.60	& 46.98	& 342.29	& 60.63 \\
& Std w/o VT, NH	& 1.30	& 13.37	& 8.58	& 7.43	& 266.68	& 11.06 \\
& \textbf{Median}	& \textbf{2.19}	& \textbf{41.72}	& \textbf{14.55}	& \textbf{47.55}	& \textbf{304.07}	& \textbf{61.14} \\
& IQR	& 1.41	& 16.28	& 13.02	& 10.46	& 328.46	& 10.84 \\

\bottomrule
\end{tabular}
\ifpreprint
}
\fi
\end{table}

\secondReviewChanges{
\section{Analysis of Treatment Budget Allocation Trends}\label{trends}
We allow the treatment budget allocations to be dynamic throughout the time horizon, which results in allocation trends. The recommended quarterly treatment budget allocation changes over time for some counties within states or remains at the same value throughout the time horizon for others. Figure~\ref{fig:bud_trends} shows the trends for the select counties that have changes in their budgets for New York and New Jersey. We chose to show New York and New Jersey because they have a visually manageable number of counties that change. Since our MIPs allocate the full budget within each period, we see that most of the counties have slightly decreasing trends in their budget allocations over time, which allows one or two counties to increase their budget allocation over time. Within our MIPs, we have a constraint ensuring that there cannot be a large decrease in the treatment budget between periods, which is why we only see slight decreasing trends. 

\begin{figure}[htbp]
     \centering
     \begin{subfigure}[b]{0.495\textwidth}
        \centering
         \includegraphics[width=\textwidth, trim={7.5cm 1cm 4.8cm 2.5cm},clip]{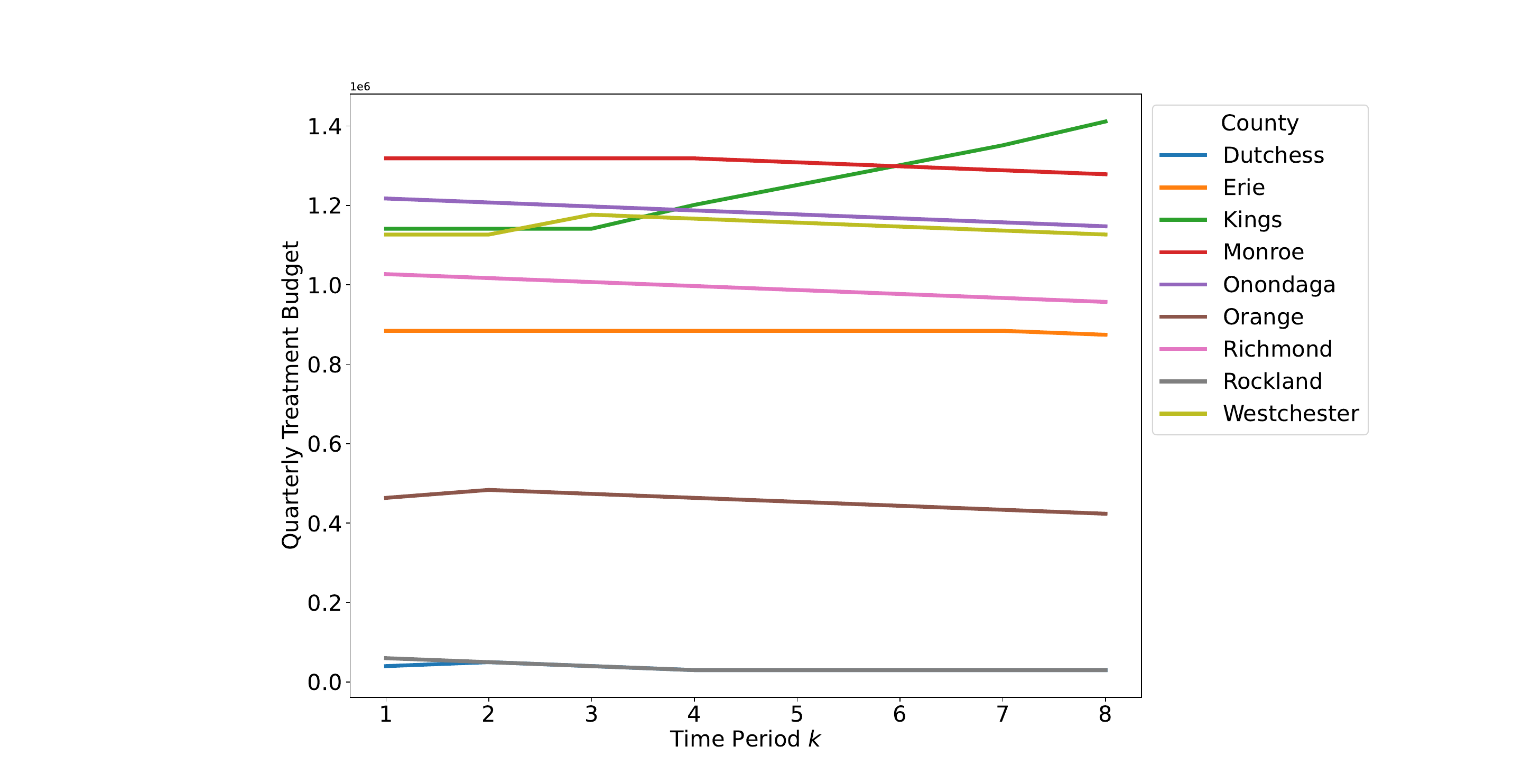}
         \caption{NY}
         \label{fig:NY_trend}
     \end{subfigure} 
        \hfill
     \begin{subfigure}[b]{0.495\textwidth}
         \centering
         \includegraphics[width=\textwidth, trim={7.5cm 1cm 4.8cm 2.5cm},clip]{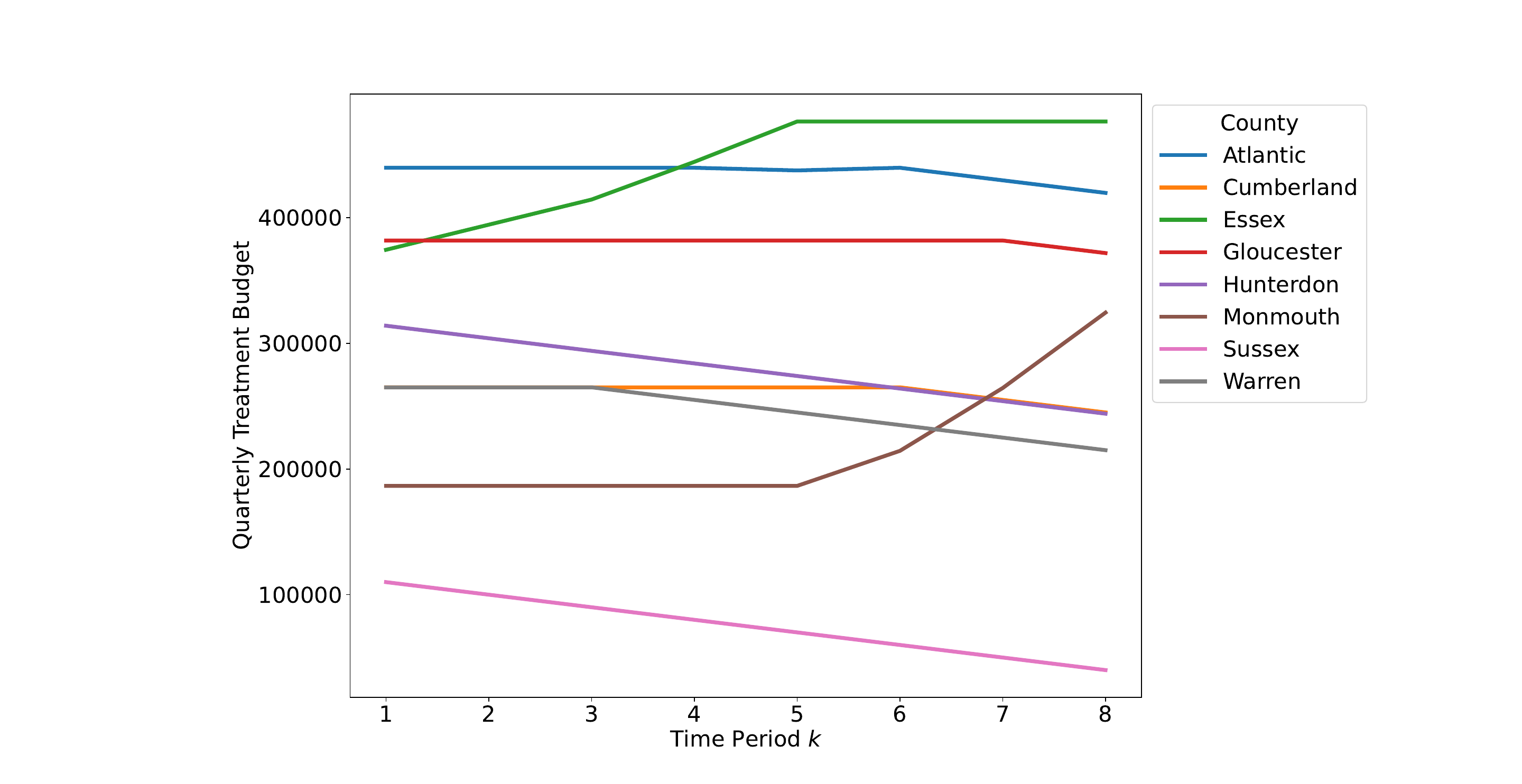}
         \caption{NJ}
         \label{fig:NJ_trend}
     \end{subfigure}
        \caption{Quarterly treatment budget allocation trends from the New York and New Jersey MIP solutions.}
        \label{fig:bud_trends}
\end{figure}

For New York, the allocated budget for Kings County increases the most over time, while for New Jersey, the allocated budget for Essex County increases before leveling off and then the Monmouth County allocated budget increases. These counties that have increases in the allocated budgets over time tend to have a larger number of treatment facilities compared to other counties in the state, while the counties that have slight decreases have a moderate number of treatment facilities. As mentioned in the main text, the MIP initially allocates a larger treatment budget to the counties with a moderate number of facilities, as these counties could benefit more from the additional treatment capacity compared to the counties with more facilities. The trends further substantiate this interpretation, as the best solution is to then slightly decrease the budget allocations to these moderate facility counties (as they start out with a larger portion of the budget), then redistribute back to the counties with more facilities that might still need more treatment capacity. 

}

\reviewChanges{
\section{Over-budget Solutions}\label{overbud}

We identify whether the solution for a state is over-budget by looking at the value of $h = \max\{\sum_{i = 1}^C x_i - N, 0\}$, which indicates how many more treatment facilities in addition to $N$ were necessary for the MIP to yield a solution. If $h > 0$, the solution is over-budget, and if $h = 0$, the solution stays within budget. 
\secondReviewChanges{
\begin{table}[htbp]
\ifpreprint \else 
\TABLEsmallX
\fi
\caption{Number of facilities over budget ($\boldsymbol h$) for states with over-budget solutions.} \label{tab:overbud}
\ifpreprint
\centering
\adjustbox{max width=0.9\textwidth}{%
\fi
\centering 
\begin{tabular}{ccccccccccccccccccc}
\toprule
\textbf{State} &\hspace{1em} AL & AR & GA & IA & KS & KY & LA & MN & MO & NE & NM & NV & OK & SC & TN & TX & VA & WI\\
\midrule
$\boldsymbol h$ &\hspace{1em} 22 & 27 & 63 & 61 & 66 & 12 & 14 & 32 & 7 & 77 & 1 & 3 & 25 & 3 & 17 & 130 & 40 & 2 \\
\bottomrule
\end{tabular}
\ifpreprint
}
\fi
\end{table}
}
Table~\ref{tab:overbud} shows the value of $h$ for each state whose MIP yields an over-budget solution. In particular, Texas, Nebraska, Georgia, Kansas, and Iowa need to obtain significantly more funding at the state level for opioid treatment facility expansion. Texas's solution is the most over budget, allocating 130 more facilities than the state has the budget for. 

This is because states with a larger number of counties tend to have more counties that currently have 0 treatment facilities. The solver needs to allocate at least 1 treatment facility to every county, which causes the solution to be over-budget. 
\thirdReviewChanges{Therefore, these over-budget solutions tend to recommend that most counties should open 1 additional treatment facility.} 


}

\thirdReviewChanges{
\section{MIP with Spatial Information}\label{spatial}
The MIP formulation with spatial information (\ie, spatial formulation) requires an additional dataset from the US Census Bureau \citep{adjacency}. The dataset indicates for each county $i$, the adjacent counties and their corresponding FIPS codes (which we call set $G_i$). We create an indicator $\mathrm{adj}_i$ for each county $i$, which is 1 if $n_i = 0$ and either all the counties or at least 3 counties
adjacent to $i$ ($j \in G_i$) have at least one initial facility open $(n_j \geq 1)$, and 0 otherwise. We only consider surrounding counties within a state, as we assume that patients would rather not travel outside the state. This gives a key indicator for whether a county $i$ with no initial facilities has surrounding counties that have facilities. 
In this case, we assume that people in that county can use the resources from the surrounding counties and, therefore, that county does not need a facility. 
Overall, we make the following modifications to our original formulation. We redefine the budget per facility as $\bar{d}_{ik}/(x_i + \mathrm{adj}_i)$, instead of $\bar{d}_{ik}/x_i$. This inclusion of $\mathrm{adj}_i$ in the denominator allows $x_i$ to be 0 in some cases. 
If $\mathrm{adj}_i = 0$, we have our original budget per facility expression. If $\mathrm{adj}_i = 1$, we make a key assumption that $x_i = 0$. This means that if the county we are considering has no facilities and the surrounding counties have facilities, there is no need to allocate any facilities or treatment budget to the county we are considering. 
We believe this is a reasonable assumption because the distance to a neighboring county is often small enough to allow patients to use a neighboring county's resources.
Enforcing this assumption requires additional constraints: $x_i \leq N(1 - \mathrm{adj}_i), \:\:\forall i \in \mathcal{C}.$ 
We also modify the lower bound of the budget per facility so it is equal to $d_\mathrm{min}$ if $\mathrm{adj}_i = 0$, and 0 if $\mathrm{adj}_i = 1$. Finally, we modify the lower bound of $\bar{d}_{ik}$ so that if $\mathrm{adj}_i = 1$ then $\mathrm{lb}_{ik} = 0$, else if $n_i = 0$ and $\mathrm{adj}_i = 0$ then $\mathrm{lb}_{ik} = d_\mathrm{min}$, else $\mathrm{lb}_{ik} = d_\mathrm{min}n_i$. Since the spatial formulation is more complex, Gurobi takes even longer to close the MIP gap when solving each state MIP. We therefore use our McCormick inequality solution method to obtain high-quality solutions.

\begin{figure}[htb]
\centering
\includegraphics[width=.6\textwidth]{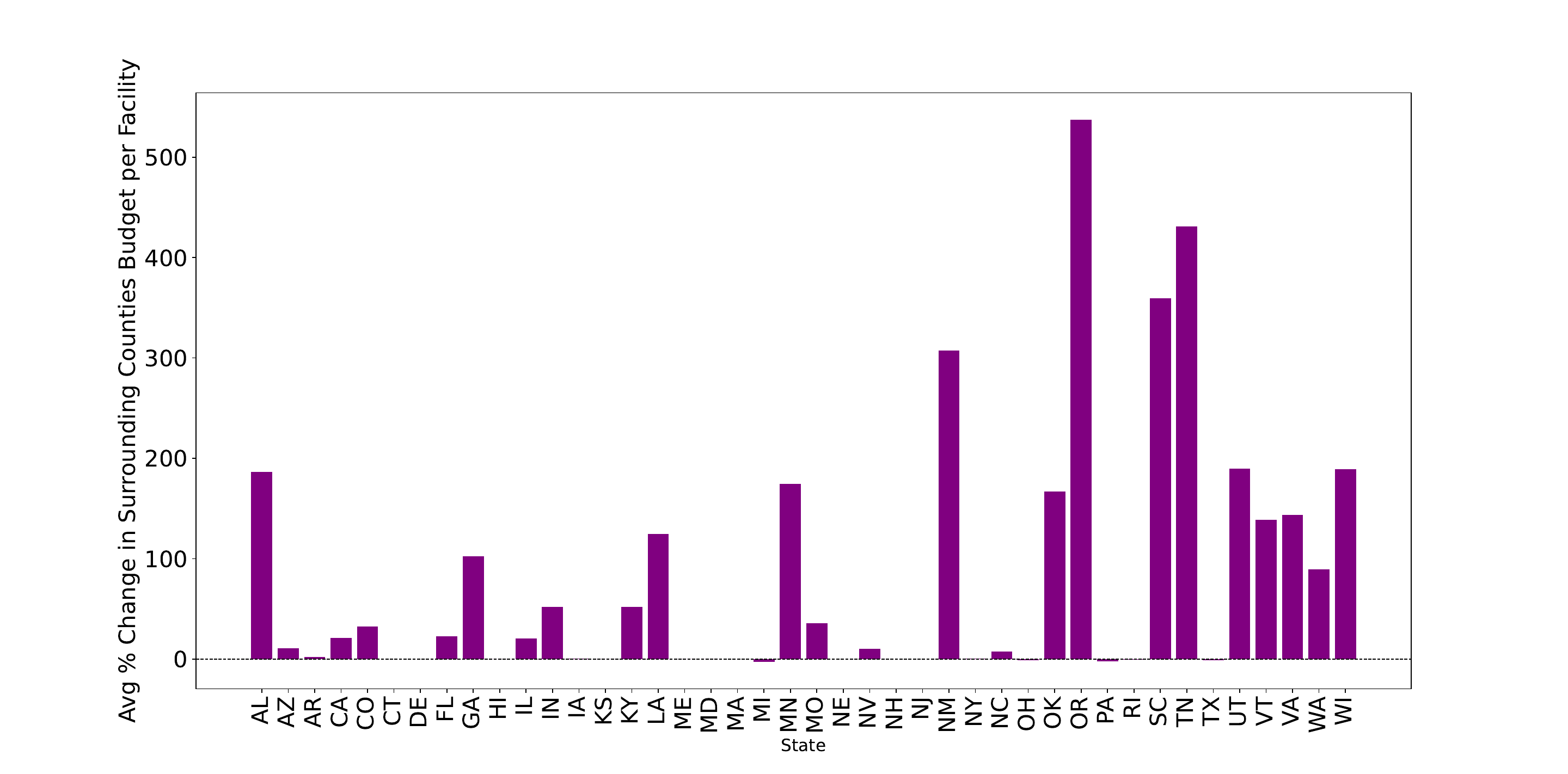}
\caption{Comparison of budget allocation per facility between spatial and original non-spatial solution. Average percentage change for counties surrounding other counties with no facilities. }
\label{fig:surr_change}
\end{figure}

The solutions to our spatial formulation show that the treatment budgets that would have been distributed to counties with no initial facilities in the original non-spatial formulation solutions, tend to, instead, be distributed to the neighboring counties. This means these neighboring counties now have additional capacity to treat patients from other counties that do not have facilities, in contrast to the solutions with no spatial information. 
Figure~\ref{fig:surr_change} shows that, in the majority of states, there is a noticeable average budget per facility increase in the surrounding counties when contrasting the spatial formulation solution with the original solution. The states that have $0\%$ bars have facilities in every county. In addition, the spatial formulation solutions are more budget savvy. As shown from Table~\ref{tab:overbud} and~\ref{tab:overbud_sp}, state solutions that are over-budget for the non-spatial formulations are either no longer over-budget for the spatial formulation or require \fourthReviewChanges{fewer} facilities (since some counties can have 0 facilities). Therefore, incorporating spatial information makes our solutions more cost-effective and allows facilities to be opened in areas where they are needed.
\begin{table}[htbp]
\ifpreprint \else 
\TABLEsmallX
\fi
\caption{Number of facilities over budget ($\boldsymbol h$) for states with over-budget solutions for spatial formulation.} \label{tab:overbud_sp}
\ifpreprint
\centering
\adjustbox{max width=0.6\textwidth}{%
\fi
\begin{tabular}{ccccccccccccccccccc}
\toprule
\textbf{State} &\hspace{1em} AL & AR & GA & IA & KS & LA & MN & NE & OK & TX & VA\\
\midrule
$\boldsymbol h$ &\hspace{1em} 11 & 14 & 48 & 55 & 62 & 6 & 16 & 77 & 14 & 120 & 14 \\
\bottomrule
\end{tabular}
\ifpreprint
}
\fi
\end{table}
}

\secondReviewChanges{
\thirdReviewChanges{
\section{Analyzing and Ensuring Robustness of MIP Solutions}\label{robust}
\subsection{Robustness Analysis of Original MIP Solutions}\label{robust1}
The robustness of the facility and budget allocation solutions vary slightly across states. 
For states like Maine and Florida, the solutions remain the same, but for states like Ohio, California, Pennsylvania, and New York, the solutions change minimally throughout the range of parameter perturbations for $\mu$ and $\zeta$, but not $\epsilon$. This is likely because $\zeta$ and $\mu$
are more closely related to the number of people in rehabilitation, which is the compartment
our decision variables are affecting.} 
\begin{figure}[htb]
     \centering
     \begin{subfigure}[b]{0.47\textwidth}
        \centering
         \includegraphics[width=\textwidth, trim={6cm 0.5 0 1cm},clip]{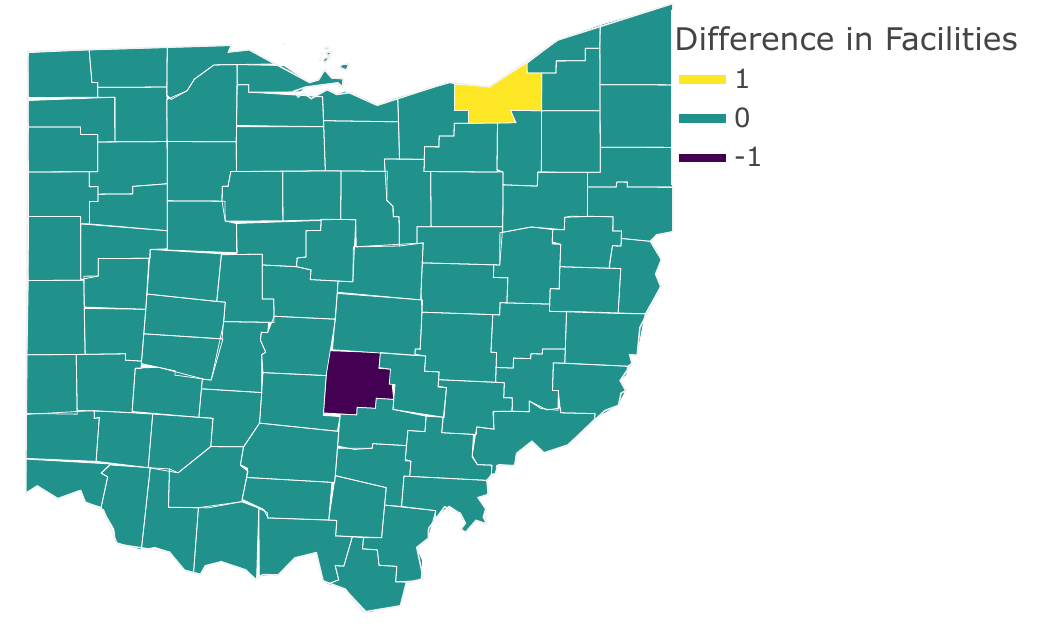}
         \caption{$\mu$ Facilities Allocation Difference}
         \label{fig:OH_mu_x_diff}
     \end{subfigure} 
        \hfill
     \begin{subfigure}[b]{0.47\textwidth}
         \centering
         \includegraphics[width=\textwidth, trim={6cm 0.5 0 1cm},clip]{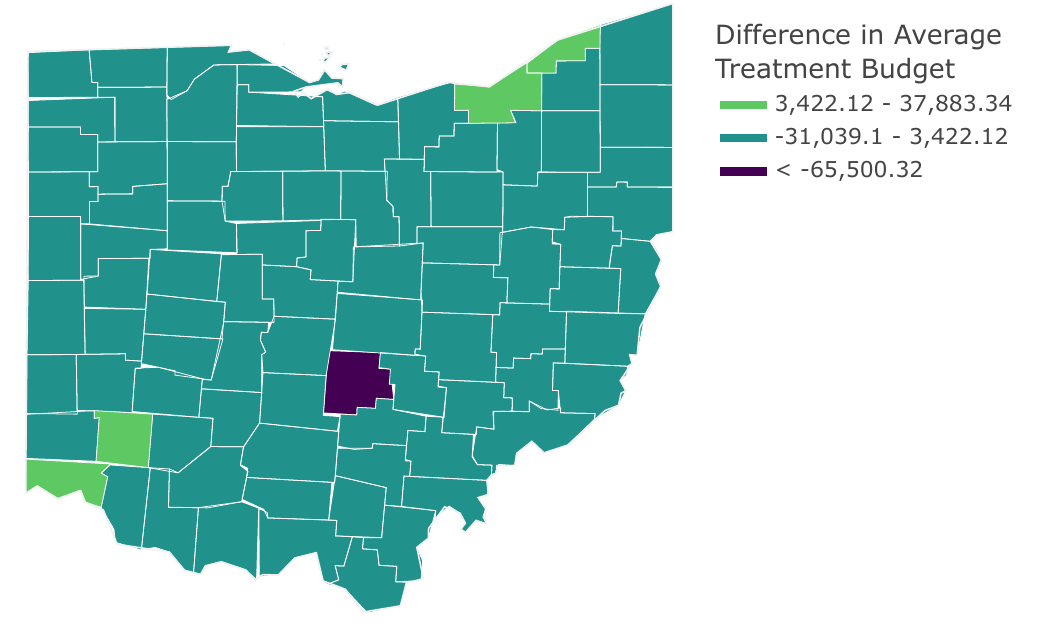}
         \caption{$\mu$ Average Budget Allocation Difference}
         \label{fig:OH_mu_bud_diff}
     \end{subfigure}
        \hfill
     \centering
     \begin{subfigure}[b]{0.47\textwidth}
        \centering
         \includegraphics[width=\textwidth, trim={6cm 0.5 0 1cm},clip]{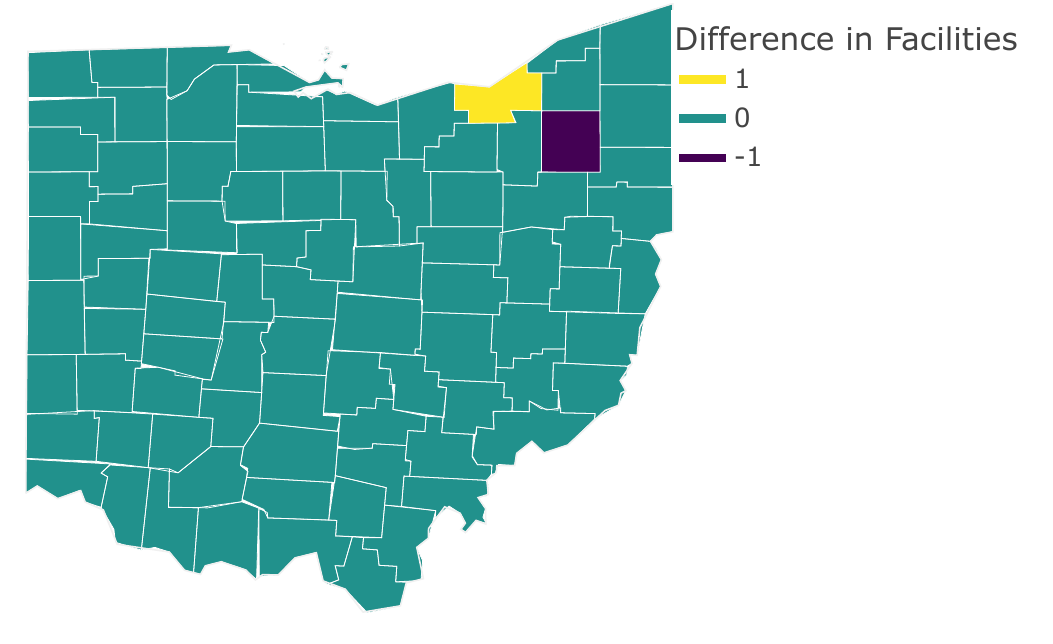}
         \caption{$\zeta$ Facilities Allocation Difference}
         \label{fig:OH_zeta_x_diff}
     \end{subfigure} 
        \hfill
     \begin{subfigure}[b]{0.47\textwidth}
         \centering
         \includegraphics[width=\textwidth, trim={6cm 0.5 0 1cm},clip]{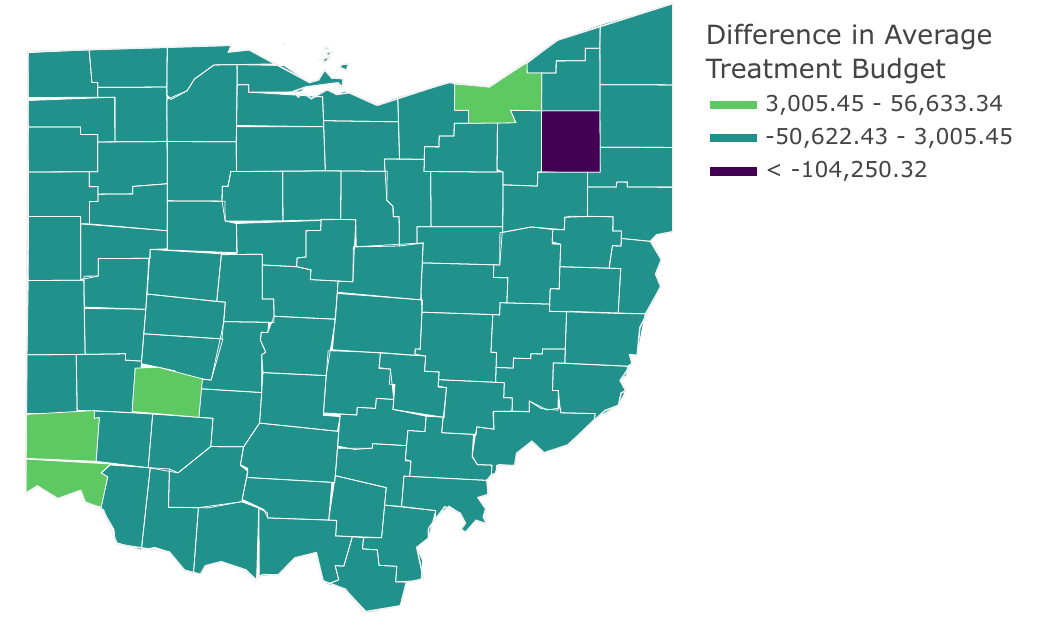}
         \caption{$\zeta$ Average Budget Allocation Difference}
         \label{fig:OH_zeta_bud_diff}
     \end{subfigure}
        \caption{Differences in Ohio MIP solutions for -80\% and 80\% deviation from the estimated $\mu$ and $\zeta$ parameters (80\% solution minus -80\% solution).}
        \label{fig:OH_diffs}
\end{figure}

Figures~\ref{fig:OH_mu_x_diff} and~\ref{fig:OH_mu_bud_diff} show the difference between the 80\% perturbed and -80\% perturbed Ohio MIP solutions for $\mu$. The facility allocation only changes minimally for 2 counties, while the budget allocation changes for 4 counties.
The differences indicate that if $\mu$ (death rate of addicts) is larger than estimated, a small amount of resources should be shifted from Fairfield to Cuyahoga. Fairfield is a small county with a low SVI ranking, while Cuyahoga is larger county with a high SVI ranking. This substantiates our observation that the MIPs allocate facilities to counties with larger populations and higher SVI rankings to minimize deaths. Figures~\ref{fig:OH_zeta_x_diff} and~\ref{fig:OH_zeta_bud_diff} further confirm this by showing similarly minimal differences in the 80\% perturbed and -80\% perturbed solutions for $\zeta$. 
These results indicate that our solutions are highly robust to parameter perturbations. 
}

\thirdReviewChanges{
\subsection{Robust MIP Formulation}\label{robust2}
We develop a robust formulation inspired by \citet{robust} to manage uncertainty resulting from our parameter estimation procedure and the population compartment data:
\begin{equation*}
\begin{array}{ll}
        \text{minimize} & e + f + \lambda_{\mathrm{pr}}\mathbf{c}_{\mathrm{pr}}^{\top}\mathbf{x} + \lambda_{\mathrm{SVI}}\mathbf{c}_{\mathrm{SVI}}^{\top}\mathbf{x}  + \lambda_{\mathrm{Pop}}\sum_{k = 0}^K \mathbf{c}_{\mathrm{Pop}}^{\top}\mathbf{\bar{d}}_{k} +\lambda_{\mathrm{inf}}\max\left(0, \sum_{i = 1}^C x_i - N\right) \\
	    \text{subject to} & \lambda_D (\mathbf{e}_K^{\top}(\bar{\mathbf{D}} + P_D \mathbf{z}_D)) \leq e, \quad \forall \mathbf{z}_D \in \mathcal{Z}_D \\
     & \lambda_A(\mathds{1}^{\top}(\bar{\mathbf{A}} + P_A\mathbf{z}_A)) \leq f, \quad \forall \mathbf{z}_A \in \mathcal{Z}_A\\
     & \text{unaffected operational constraints}\\
		& {\rm\bf z}_{k+1} = \bar{f}({\rm\bf z}_k, {\rm\bf x}, \bar{{\rm\bf d}}_k), \quad k=0,\dots,K-1\\
        & -\mathbf{e}_k^{\top}\bar{\mathbf{V}} + \mathbf{e}_k^{\top}P_V \mathbf{z}_V \leq 0, \quad \forall\: V \in \{S, P, I, A, R, D\}, \mathbf{z}_V \in \mathcal{Z}_V, k \in \mathcal{K}\\
         & \text{unaffected domain constraints}
        \end{array}
\end{equation*}
\fourthReviewChanges{We define $\bar{\mathbf{V}} \in \{\bar{\mathbf{S}}, \bar{\mathbf{P}}, \bar{\mathbf{I}} , \bar{\mathbf{A}}, \bar{\mathbf{R}}, \bar{\mathbf{D}}\}$ as vectors in $\mathbb{R}^{K+1}$ that represent the nominal population compartment values from our dynamical model $\bar{f}$. The dynamics $\bar{f}$ remain the same as in our original formulation, and give the nominal compartment values $\bar{V}_k$ at each time $k = 1, \dots, K$. As \citet{robust}, we robustify the constraints that involve these nominal compartment values}. The uncertainty sets are of the form: $\mathcal{Z}_V = \{\mathbf{z}_V \in \mathbb{R}^{K+1}: \|\mathbf{z}_V\|_{\infty} \leq \rho_V \}, \:\: \forall\: V \in \{S, P, I, A, R, D\}.$
We define $\mathbf{e}_k$ as the unit vector where the $k$th element is 1. We also define matrices for each compartment which allow us to increase the size of the uncertainty set at each time period to account for less accurate predictions in the far future. We consider that uncertainty increases over time, by defining $P_V \in \mathbb{R}^{(K+1)\times (K+1)}$ as follows:
$$P_V = {\bf diag}(\psi^0\bar{V}_0,\dots, \psi^K\bar{V}_K), \quad \forall\: V \in \{S, P, I, A, R, D\},$$
where $\psi > 1$.  We can then obtain the robust counterpart of the constraints with standard robust optimization techniques. 
We choose $\rho_S = \rho_P =\rho_I = \rho_A= \rho_R = \rho_D = 0.4$ and coefficient $\psi = 1.1$. 
A $\psi$ value of 1.1 indicates that we assume the uncertainty is growing by 10\% after each time period (\ie, we are 10\% less sure about the values of these compartments with respect to the nominal value after each time period). A $\rho$ value of 0.4 indicates that we allow up to 40\% deviation in either direction compared to the nominal compartment value. 
}

\thirdReviewChanges{
Table~\ref{tab:add_robust} shows that for most states, the robust solution leads to additional benefits (compared to the non-robust solution) in terms of the percentage changes in the values of $A$, $R$, and $D$ in comparison to the baseline compartmental model dynamics. The following states are not included in the table as the robust solutions are the same as the original solutions: Delaware, Iowa, Kansas, Maine, Nebraska, New Hampshire, Rhode Island, and Vermont. This indicates that for these states, the original solutions are highly robust to deviations in the population compartment values. A small number of states (Maryland, Massachusetts, New Jersey, Ohio) have slightly smaller percentage changes compared to the non-robust solutions, but the differences are minor. 
Overall, the robust solutions behave as expected, because the value of the components of the objective related to the $A$ and $D$ compartments now take into account worst-case realizations of the uncertainty. Therefore, the robust problem encourages solutions that achieve lower nominal objectives compared to our original formulation, which, on average, brings beneficial variations in the compartment values $A, R,$ and $D$.
\begin{table*}[ht]
\centering
\setlength\tabcolsep{0.15em}
\footnotesize
  \begin{minipage}{1\textwidth}
  \centering
  \caption{Additional percentage point change in the $A$, $R$, $D$ values resulting from the robust versus the non-robust formulation solutions, in comparison to baseline dynamics.}
  \label{tab:add_robust}
  \adjustbox{max totalheight=0.1\textheight}{
  \begin{tabular*}{\textwidth}{@{\extracolsep{\fill}}ccccccccccccccccccc@{\extracolsep{\fill}}}
    \toprule
    & AL &	AZ &	AR &	CA & CO  &	CT &	FL &	GA	& HI & IL	& IN & KY & LA & MD &	MA	& MI & MN & MO\\
    \midrule
    \begin{filecontents*}{./Data/additional_percent_changes_no0.csv}
Compartment,AL,AZ,AR,CA,CO,CT,FL,GA,HI,IL,IN,KY,LA,MD,MA,MI,MN,MO,NV,NJ,NM,NY,NC,OH,OK,OR,PA,SC,TN,TX,UT,VA,WA,WI,Mean,Std
A,-0.04,-0.34,-0.16,-0.62,-0.05,-0.03,-0.47,-2.57,-0.85,0,-0.24,-1.24,-0.23,0.02,0.01,-0.86,-0.27,-0.2,-0.43,0.13,-0.35,-0.01,-0.42,0.08,-0.18,-0.03,-0.49,-0.06,-2.83,-1.59,-0.09,-1.21,-0.09,-0.62,-0.48,0.69
R,0.35,3.32,5.25,3.8,1.21,0.07,4.73,21.52,6.78,0.02,2.21,18.14,3.04,-0.05,-0.03,7.16,1.75,2.89,8.32,-0.67,1.27,0.02,2.7,-1,1.9,0.3,2.5,0.69,36.73,24.25,0.91,11.84,0.8,6.59,5.27,8.27
D,0,-0.01,0,-0.03,0,0,-0.03,-0.09,-0.05,0,-0.01,-0.06,0,0,0,-0.05,0,-0.01,-0.02,0.01,0,0,-0.02,0.01,0,0,-0.03,0,-0.14,-0.09,0,-0.09,0,-0.01,-0.02,0.04
\end{filecontents*}
\csvreader[head to column names, late after line=\\]{./Data/additional_percent_changes_no0.csv}{
    Compartment = \cb,
    AL = \a, AZ = \b, AR = \c,	CA = \d, CO = \e, CT = \f, FL = \g,	GA = \h, HI = \i,  IL = \j, IN = \k, KY = \l,  LA= \m, MD=\n,	MA	= \o, MI =\p, MN = \q, MO = \r
    }{\cb & \a & \b & \c & \d & \e & \f & \g & \h & \i & \j & \k & \l & \m & \n & \o & \p & \q & \r}
    \bottomrule
  \end{tabular*}
  }
  \end{minipage}
    \hfill
    \begin{minipage}{\textwidth}
    \centering
  \adjustbox{max totalheight=0.1\textheight}{
  \begin{tabular*}{\textwidth}{@{\extracolsep{\fill}}ccccccccccccccccccc@{\extracolsep{\fill}}}
    \toprule
    & NV & NJ & NM &	NY	& NC & OH & OK	& OR	& PA	& SC & TN &	TX	& UT & VA	& WA &	WI & \textbf{Mean} & \textbf{Std}\\
    \midrule
    \csvreader[head to column names, late after line=\\]{./Data/additional_percent_changes_no0.csv}{
    Compartment = \cb, NV =\s, NJ =\t, NM= \u,	NY =\v, NC = \w,  OH = \x, OK = \y,	OR	= \z, PA = \aa, SC = \bb,  TN = \cc,	TX	= \dd, UT = \ee,  VA = \ff,  WA  = \gg, WI = \hh, Mean = \mean, Std = \std
    }{\cb & \s & \t & \u & \v & \w & \x & \y & \z & \aa & \bb & \cc &\dd & \ee & \ff & \gg & \hh & \mean & \std}
    \bottomrule
  \end{tabular*}
  }
    \end{minipage}
    
    {\vspace{0.5em}\footnotesize Note: We do not include states that had no changes in the compartment values.}
\end{table*}
}
\secondReviewChanges{
\section{Recommended Facility Location Solution Comparison with Benchmarks}\label{compbench}

Figure~\ref{fig:comp_bench} shows a comparison between our optimized facility location solution for Florida and demographic-based benchmarks. \thirdReviewChanges{We show Florida, as it is a sufficiently geographically diverse state in terms of population and SVI. This gives a characteristic example of how the benchmarks differ from each other, and especially how they differ from our optimized solution for a state.} 

\begin{figure}[ht]
     \centering
     \begin{subfigure}[b]{0.49\textwidth}
        \centering
         \includegraphics[width=\textwidth, trim={6cm 0.5 0 1cm},clip]{./Figures/map_x_FL.pdf}
         \caption{Optimized}
         \label{fig:opt}
     \end{subfigure} 
        \hfill
     \begin{subfigure}[b]{0.49\textwidth}
         \centering
         \includegraphics[width=\textwidth, trim={6cm 0.5 0 1cm},clip]{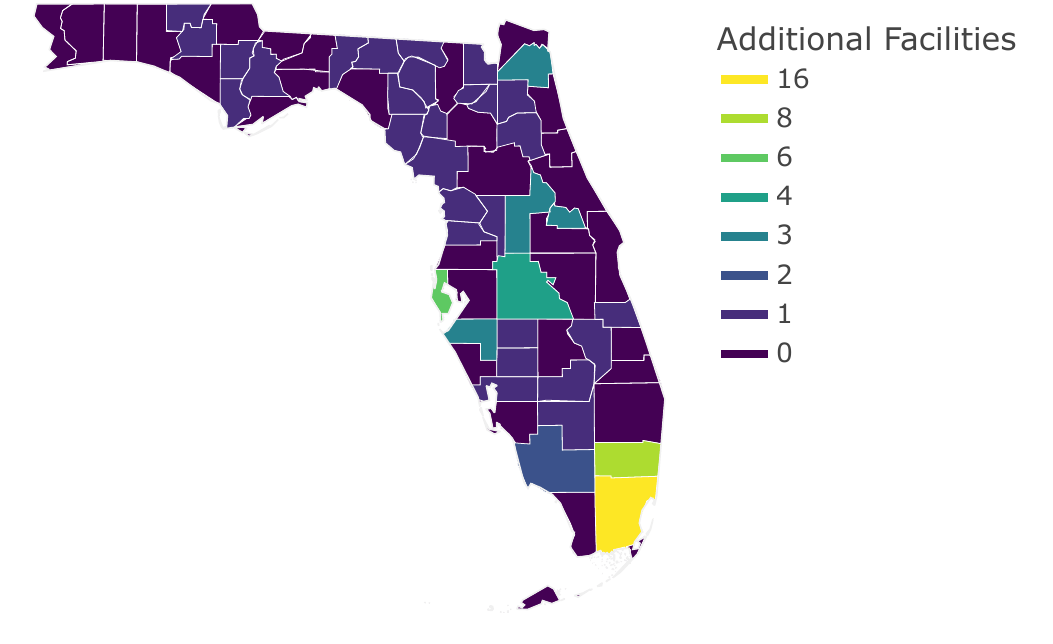}
         \caption{Population-Based}
         \label{fig:pop}
     \end{subfigure}\\[1ex]
     \begin{subfigure}[b]{0.49\textwidth}
        \centering
         \includegraphics[width=\textwidth, trim={6cm 0.5 0 1cm},clip]{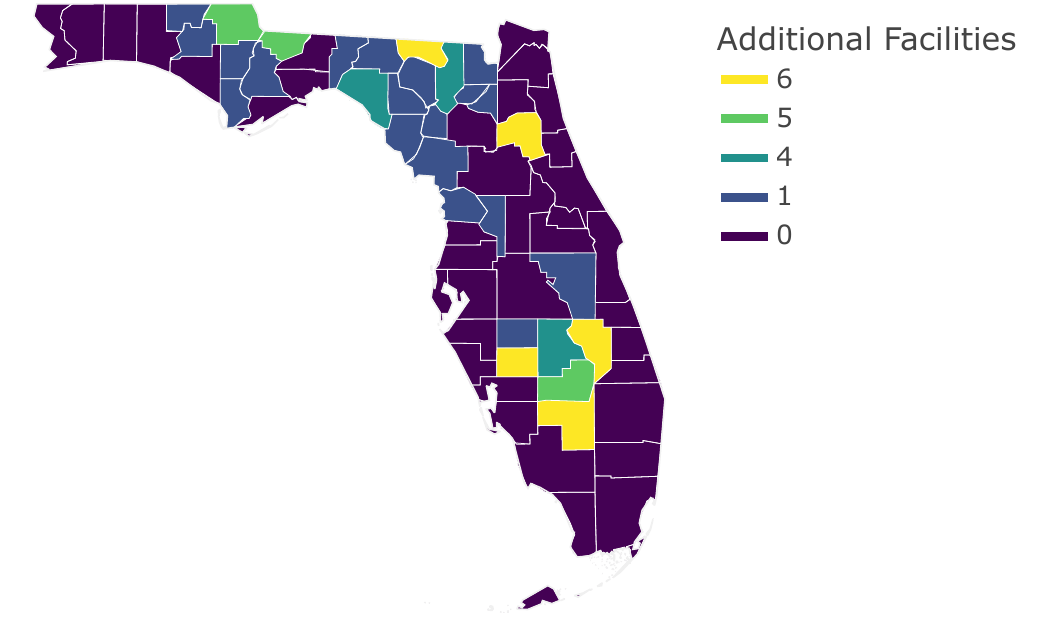}
         \caption{SVI Ranking-Based}
         \label{fig:svi}
     \end{subfigure} 
        \hfill
    \begin{subfigure}{0.49\textwidth}
        \centering
        \includegraphics[width=\textwidth, trim={6cm 0.5 0 1cm},clip]{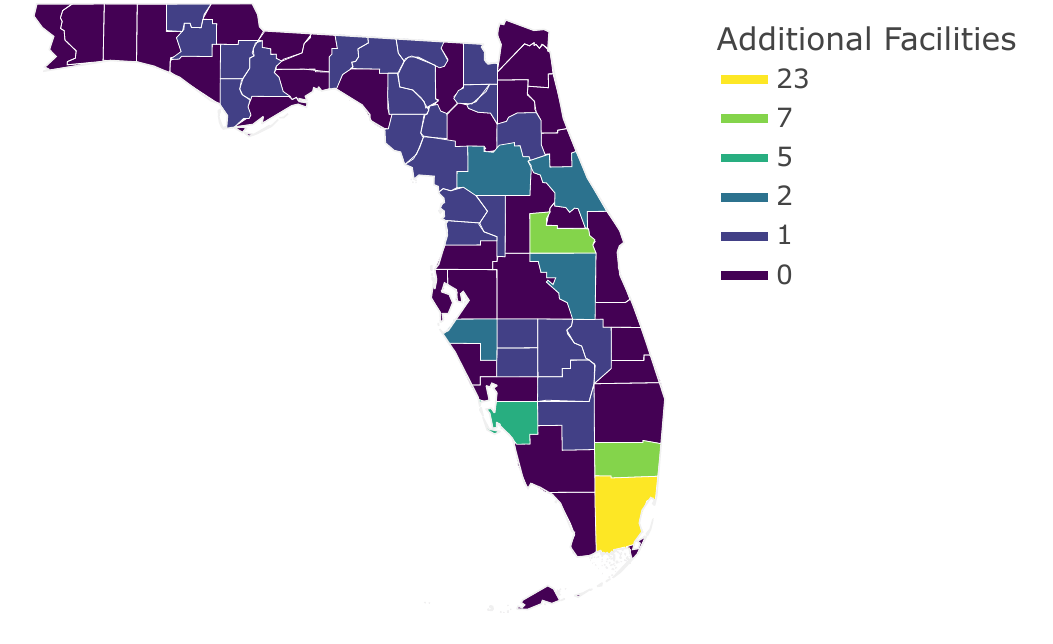}
        \caption{Population-weighted SVI-Based}
        \label{fig:psvi}
    \end{subfigure}
        \caption{Comparing optimized facility location solutions for Florida with benchmarks.}
        \label{fig:comp_bench}
\end{figure}
\noindent We compare to the population-based and SVI ranking-based benchmarks shown in the main text, as well as a population-weighted SVI (SVI ranking times county population) benchmark. The optimization problem for the population-weighted SVI benchmark is the same as the other two benchmarks, just using different data. This benchmark considers both SVI ranking and county population. Our optimized solution is quite different from just allocating resources proportional to population, SVI ranking, or population-weighted SVI. Our solution is more balanced in terms of the way it allocates additional facilities. There is not a single county that gets significantly more additional facilities than all other counties, which we can see occurs within the population and population-weighted SVI benchmarks. However, our solution does allocate facilities to counties that have higher populations (whereas the SVI-based solution does not), in addition to allocating to counties that have high SVI rankings (even though it does not allocate as many facilities to counties with high SVI rankings).
}

\thirdReviewChanges{
\section{Optimality Gap and Solve Time}\label{opt_gap}
Since our solution method is a relaxation, we provide optimality gaps, \secondReviewChanges{which we define as the gap between the upper and lower bounds on the objective obtained from our
solution method}, for the original formulation 
(full table in \ifpreprint Appendix~\ref{tab_fig}). \else Section~\ref{tabfigs}). \fi The median of the optimality gaps is 4.28\% (mean = 3.99\%), with a maximum of 10.39\%.
This means that the feasible solutions we obtain are very close to optimal, and in some cases are the optimal solution. 
\secondReviewChanges{For instance, the optimality gaps for Iowa, Kansas, and Nebraska are essentially 0 (less than $10^{-7}$),
and we confirm that the feasible facility allocations we obtain from our method are the same as the optimal solutions.} 
This shows that we obtain a strong relaxation from our method.
\begin{figure}[htb]
\centering
\includegraphics[width=.8\textwidth,trim={0 0 1cm 1cm},clip]{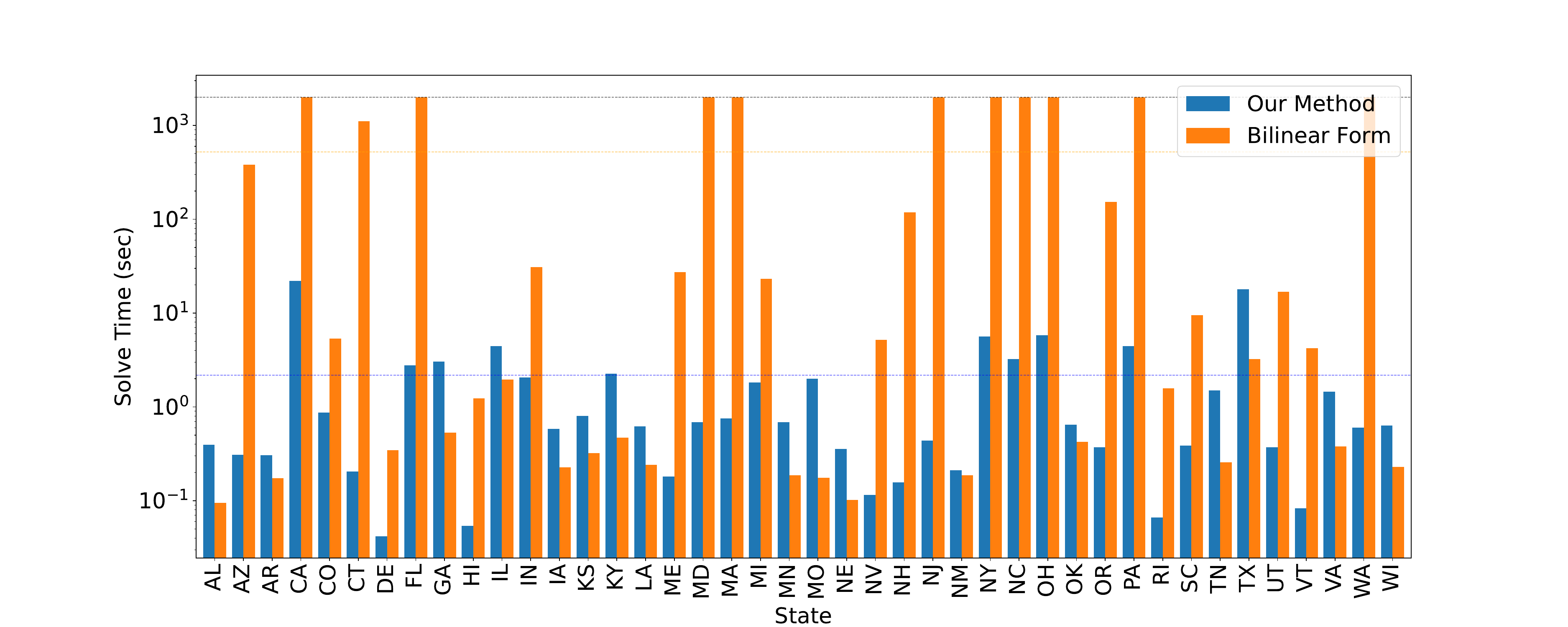}
\caption{Solve time comparison between our relaxed formulation and the original bilinear formulation. }
\label{fig:solve_times}
\end{figure}

In addition, our solution method is more efficient than solving the bilinear problem in Gurobi.
\secondReviewChanges{The average solve time for our method is 2.179 seconds (blue dashed line) compared to the average solve time for the original bilinear formulation which is 521.220 seconds (orange dashed line). Gurobi is also unable to converge within the time limit of 2000 seconds for larger bilinear problem instances. We show the MIP gaps for the bilinear problem instances that cannot be solved within the time limit in Table~\ref{tab:mipgap}. }
\begin{table}[htbp]
\ifpreprint \else 
\TABLEsmallX
\fi
\caption{MIP gaps (in \%) for bilinear problem instances that could not be solved within the time limit.} \label{tab:mipgap}
\ifpreprint
\centering
\adjustbox{max width=0.7\textwidth}{%
\fi
\centering 
\begin{tabular}{ccccccccccc}
\toprule
\textbf{State} &\hspace{1em} CA & FL & MA & MD & NC & NJ & NY & OH & PA & WA\\
\midrule
\textbf{MIP Gap} &\hspace{1em} 1.58 & 1.45 & 0.51 & 0.88 & 0.21 & 0.49 & 1.67 & 0.81 & 0.92 & 0.51 \\
\bottomrule
\end{tabular}
\ifpreprint
}
\fi
\end{table}
\reviewChanges{
Figure~\ref{fig:solve_times} shows that for all bilinear MIPs that Gurobi was unable to solve to optimality within the time limit (indicated by the black dashed line), our relaxed solution method provides high-quality solutions in an order of magnitude less time. Although Gurobi solves certain bilinear MIPs faster, the order of magnitude difference is not as significant. 
}
}

\section{Tables and Figures}\label{tab_fig}

Table~\ref{tab:state_fit_params} shows the exact values of the parameters we estimate using the \secondReviewChanges{neural ODE-inspired model fitting process} for each state.  \secondReviewChanges{Table~\ref{tab:change} shows the effect of optimizing the locations of additional treatment facilities and the treatment budget on the compartments $A$, $R$, and $D$ after 2 years for almost all US states, and the optimality gaps (gaps between the upper and lower bounds of our solution method). \thirdReviewChanges{Table~\ref{tab:eqm} shows the equitability loss for all state benchmarks and optimized solutions.}
}
\begin{table*}[htb]
    \small
  \centering
  \caption{Estimated parameters for each US state.}
  \label{tab:state_fit_params} 
  \adjustbox{max totalheight=0.9\textheight, center}{%
  \begin{tabular*}{\textwidth}{@{\extracolsep{\fill}}llllll@{\extracolsep{\fill}}}
    \toprule
    State &   $\phi$   &  $\epsilon$ & $\beta$ & $\zeta$ & $\mu$\\
    \midrule
    \begin{filecontents*}{./Data/avg_weighting_params_display.csv}
ph,ep,be,ze,m,State
0,1.54,0.000425,0.1194,0.00247,AL
0.0020,2.459,0,0.0857,0.00422,AZ
0,2.075,0.000514,0.0298,0.00517,AR
0,3.803,0.000438,0.141,0.00593,CA
0,2.98,0.000368,0.0395,0.00452,CO
0.0625,2.77,0.006127,0.3831,0.00759,CT
0.0800,1.924,0.008459,0.201,0.00485,DE
0,2.419,0.000545,0.0864,0.00778,FL
0,2.769,0.000324,0.0981,0.00478,GA
0,4.245,0.000887,0.111,0.00826,HI
0,3.155,0.00086,0.1644,0.00674,IL
0,1.851,0.001032,0.1001,0.00394,IN
0,2.811,0.000919,0.0341,0.00536,IA
0,2.755,0.0014,0.0995,0.0068,KS
0.0362,1.423,0.001724,0.0504,0.00535,KY
0,2.157,0.000654,0.0706,0.00313,LA
0.1266,2.171,0.010594,0.294,0.00943,ME
0.0702,2.745,0.005046,0.4335,0.01031,MD
0.1333,2.988,0.011304,0.2559,0.007,MA
0,2.245,0.000599,0.1079,0.00696,MI
0,4.218,0.001198,0.1414,0.00562,MN
0,2.435,0.000466,0.0571,0.00691,MO
0,3.576,0.000469,0.05,0.00386,NE
0,2.267,0.000007,0.0482,0.00856,NV
0.0538,2.32,0.00471,0.119,0.00654,NH
0.0363,3.184,0.004651,0.1717,0.00661,NJ
0.0238,2.533,0.001564,0.2593,0.01806,NM
0.0628,3.925,0.0051,0.4042,0.00889,NY
0,2.109,0.000393,0.1334,0.00778,NC
0.0924,1.984,0.005545,0.0634,0.00814,OH
0,1.875,0.000714,0.0943,0.01446,OK
0,2.084,0.000533,0.1025,0.00618,OR
0.0286,2.364,0.003446,0.1761,0.00651,PA
0.0729,2.411,0.005088,0.4297,0.00986,RI
0,2.347,0.000084,0.08,0.00425,SC
0,1.697,0.00066,0.0644,0.00621,TN
0,4.344,0.000489,0.0579,0.00446,TX
0,2.276,0.000335,0.1019,0.00977,UT
0.1382,3.225,0.019049,0.1954,0.00636,VT
0,2.563,0.000396,0.088,0.00743,VA
0,2.355,0.000191,0.0913,0.00591,WA
0,2.647,0.001311,0.087,0.00817,WI
\end{filecontents*}
\csvreader[head to column names, late after line=\\]{./Data/avg_weighting_params_display.csv}{
    State=\state,
    ph=\p, ep=\ep, be = \b, ze = \z, m = \m
    }{\state & \p & \ep & \b & \z & \m}
    \bottomrule
  \end{tabular*}
  }
    
    \vspace{0.2em}
    {\footnotesize Note: We treated all parameters that were less than $10^{-7}$ as approximately 0.}
\end{table*}

\begin{table*}[ht]
  \centering
  \begin{minipage}{0.9\textwidth}
  \caption{Effect of our state MIP solutions on $A$, $R$, and $D$ values in comparison to baseline epidemic dynamics after 2 years and the optimality gap (in \%).}
  \label{tab:change}
  \adjustbox{max totalheight=0.9\textheight, center}{
  \begin{tabular*}{\textwidth}{@{\extracolsep{\fill}}lcccc@{\extracolsep{\fill}}}
    \toprule
    State & $A$ & $R$ & $D$ & Gap\\
    \midrule
    \begin{filecontents*}{./Data/rev2_percent_changes_gap_display.csv}
State,A,R,D,Gap
AL,-14.54,106.73,-0.77,0.0053
AR,-12.28,355.64,-0.86,0.0194
AZ,-6.68,63.91,-0.32,6.0975
CA,-9.84,60.03,-0.56,6.2421
CO,-10.69,219.64,-0.47,3.2487
CT,-3.42,7.92,-0.24,4.4892
DE,-3.55,14.72,-0.34,2.4539
FL,-4.28,42.58,-0.31,6.4298
GA,-14.82,120.77,-0.70,1.7859
HI,-8.11,64.20,-0.54,10.3912
IA,-11.01,261.05,-0.79,0
IL,-2.30,12.49,-0.16,8.0755
IN,-6.13,52.94,-0.43,4.4144
KS,-19.05,161.51,-1.13,0
KY,-2.03,28.89,-0.17,6.9904
LA,-22.00,255.96,-0.99,0.0334
MA,-4.43,14.82,-0.38,3.6756
MD,-5.72,11.31,-0.41,5.7144
ME,-2.09,6.17,-0.26,2.7943
MI,-5.01,41.55,-0.35,6.4411
MN,-18.66,105.41,-0.90,0.0388
MO,-4.56,63.63,-0.26,8.8398
NC,-3.13,19.49,-0.25,6.2854
NE,-16.99,309.34,-1.18,0
NH,-6.41,39.77,-0.45,4.5723
NJ,-3.20,16.65,-0.24,6.4980
NM,-23.10,77.17,-1.75,0.0820
NV,-10.54,200.19,-0.72,1.7389
NY,-4.75,10.70,-0.32,6.3681
OH,-5.43,68.31,-0.40,7.5116
OK,-20.07,193.69,-1.60,0.0222
OR,-7.91,66.50,-0.55,2.9939
PA,-8.07,40.79,-0.61,6.2847
RI,-14.19,28.93,-0.84,2.7134
SC,-9.44,99.52,-0.50,2.0697
TN,-6.69,87.99,-0.35,9.5162
TX,-3.27,50.05,-0.21,6.4834
UT,-8.35,72.98,-0.57,2.7482
VA,-3.73,36.73,-0.30,4.3762
VT,-14.16,57.54,-1.01,4.1843
WA,-4.95,43.62,-0.34,4.5858
WI,-13.84,135.81,-0.87,0.1899
\end{filecontents*}
\csvreader[head to column names, late after line=\\]{./Data/rev2_percent_changes_gap_display.csv}{
    State=\state,
    A=\A, R=\R, D = \D, Gap=\gap
    }{\state & \A & \R & \D & \gap}
    \bottomrule
  \end{tabular*}
  }

    \end{minipage}
\end{table*}

\begin{table*}[ht]
  \centering
  \begin{minipage}{\textwidth}
  \caption{Equitability loss measure for state benchmarks and optimized solutions.}
  \label{tab:eqm}
  \adjustbox{max totalheight=0.9\textheight, center}{
  \begin{tabular*}{\textwidth}{@{\extracolsep{\fill}}lcccccc@{\extracolsep{\fill}}}
    \toprule
    State & Pop Base & Pop Spatial & SVI Base & SVI Spatial & Opt & Opt Spatial\\
    \midrule
    \begin{filecontents*}{./Data/equitability_measure_table_spatial.csv}
State,Pop Base,Pop Spatial,SVI Base,SVI Spatial,Opt,Spatial Opt
AL,34.55,42.74,34.55,42.74,35.06,42.74
AZ,200.19,217.51,145.23,147.51,175.51,173.51
AR,30.31,37.61,30.31,37.61,30.30,37.61
CA,596.79,571.21,409.21,277.47,583.48,600.99
CO,177.60,182.40,173.91,145.38,164.31,178.01
CT,130.62,51.69,87.53,8.25,40.69,41.69
DE,28.50,30.01,26.50,10.01,26.50,26.50
FL,274.16,318.48,212.04,206.50,299.02,319.27
GA,97.39,108.28,97.39,108.28,100.12,108.28
HI,12.81,17.33,5.29,6.14,7.33,9.32
IL,350.15,340.59,345.13,306.63,316.29,341.83
IN,179.93,181.88,166.05,129.86,145.80,183.94
IA,69.33,68.64,69.33,68.64,69.75,68.64
KS,72.19,70.15,72.19,70.15,71.17,70.15
KY,174.52,213.37,174.52,179.37,170.68,212.83
LA,51.64,58.47,51.64,58.47,51.87,58.47
ME,82.93,84.68,74.28,54.68,75.86,77.81
MD,248.57,280.13,262.91,172.51,275.22,285.01
MA,235.19,183.87,287.05,49.42,129.87,129.87
MI,170.12,202.53,157.08,138.70,167.45,200.28
MN,89.02,93.80,89.02,93.80,84.30,93.80
MO,153.57,197.59,153.57,161.59,160.60,197.59
NE,77.77,77.77,77.77,77.77,77.77,77.77
NV,38.22,42.47,33.75,32.47,34.62,34.25
NH,65.96,61.69,65.04,19.69,42.97,44.62
NJ,127.24,134.34,151.42,78.73,114.34,116.34
NM,54.58,61.55,54.58,54.52,55.39,58.15
NY,688.55,658.51,576.57,448.51,676.51,680.51
NC,266.88,311.44,228.88,237.80,286.02,322.81
OH,364.11,370.31,329.07,213.91,354.33,383.29
OK,46.96,53.23,46.96,53.23,45.28,53.23
OR,95.99,114.23,78.80,72.23,89.88,95.27
PA,343.76,331.94,273.41,211.18,354.17,376.33
RI,69.07,17.66,69.60,7.60,14.06,19.46
SC,53.69,92.49,44.20,58.51,65.58,78.40
TN,116.75,142.17,116.75,142.17,125.85,142.17
TX,154.96,159.79,154.96,159.79,153.87,159.79
UT,145.18,151.00,135.68,117.70,131.07,140.97
VT,23.78,25.53,26.80,14.23,17.53,19.50
VA,108.57,119.51,108.57,119.51,109.22,119.51
WA,178.05,201.53,130.90,121.53,173.13,187.46
WI,103.42,116.68,103.42,88.45,91.22,114.50
\end{filecontents*}
\csvreader[head to column names, late after line=\\]{./Data/equitability_measure_table_spatial.csv}{
    State=\state,
    Pop Base = \popb, Pop Spatial = \pops,	SVI Base = \svib,	SVI Spatial = \svis, Opt = \opt, Spatial Opt = \opts
    }{\state & \popb & \pops & \svib & \svis & \opt & \opts}
    \bottomrule
  \end{tabular*}
  }
  \end{minipage}
\end{table*}

\clearpage

\end{appendices}
\fi

\end{document}

%% file: definitions.tex
\newcommand{\eg}{{\it e.g.}}
\newcommand{\ie}{{\it i.e.}}

\newcommand{\reals}{{\mbox{\bf R}}}
\newcommand{\integers}{{\mbox{\bf Z}}}

\newcommand{\loss}{\mathcal{L}}
